\documentclass[amsmath,amssymb,aps,pra,floatfix,twocolumn]{revtex4-1}

\usepackage{hyperref}
\usepackage{color}
\usepackage{graphicx}
\usepackage{tikz}% drawing package
\usetikzlibrary{positioning,shapes,snakes,calc}
\usepackage{blindtext}

\begin{document}

\title{Interferometric Quantum Cascade Systems}

\author{Stefano Cusumano}
\affiliation{NEST, Scuola Normale Superiore and Istituto Nanoscienze-CNR, I-56127 Pisa, Italy}
\email{stefano.cusumano@sns.it}

\author{Andrea Mari}
\affiliation{NEST, Scuola Normale Superiore and Istituto Nanoscienze-CNR, I-56127 Pisa, Italy}
\email{andrea.mari@sns.it}

\author{Vittorio Giovannetti}
\affiliation{NEST, Scuola Normale Superiore and Istituto Nanoscienze-CNR, I-56127 Pisa, Italy}
\email{vittorio.giovannetti@sns.it}

\begin{abstract}
In this work we consider quantum cascade networks in which quantum systems  are connected through unidirectional channels that can mutually interact giving rise to interference effects.
In particular we show how to compute master equations for cascade systems in an arbitrary interferometric configuration by means of a collisional model. We apply our general theory to two specific examples: the first consists in two systems arranged in a  Mach-Zender-like configuration; the second is a three system network where it is possible to tune the effective chiral interactions between the nodes exploiting interference effects.
\end{abstract}
\maketitle

\section{Introduction}

Quantum cascade systems (QCSs) describe those physical situations where 
a first party (the {\it controller}) can influence the dynamic of a second  party (the {\it idler}) without being affected by the latter.
 The  asymmetric character of these couplings  originates from the presence of  an environmental medium  (e.g. an optical isolator~\cite{NATURE} or a bosonic chiral channel) which acts  as mediator of the  interactions and which  allows for  unidirectional propagation of pulses from a controller to its associated idler. 
First interests in these models grew in the 80's because of the necessity of a formalism able to take into account  the reaction
of a quantum system (say an atom or a electromagnetic cavity) to the light emitted by another one \cite{PhysRevLett.70.2273,PhysRevA.50.1792,PhysRevA.31.3761,PhysRevLett.70.2269,gardinerbook, gard-zoll-book,Vogel2006}. In recent years there has been a revival of interest towards QCSs, due to the possibility of creating entangled states and other tasks for quantum computation \cite{PhysRevLett.113.237203,PhysRevA.89.022335,PhysRevA.84.042341,1367-2630-14-6-063014}, chiral optical networks \cite{PhysRevA.91.042116,PhysRevLett.116.093601,PhysRevA.86.042306}, and in the managing of heat transmission \cite{PhysRevA.91.022121}; 
also several experimental implementations  have been proposed, exploiting, for instance, nanophotonic waveguides \cite{Sollner:2015aa,1606.07466} and spin-orbit coupling \cite{Petersen67}.

 %%%%%%
  \begin{figure}[!t]
\centering
\includegraphics[scale=0.35]{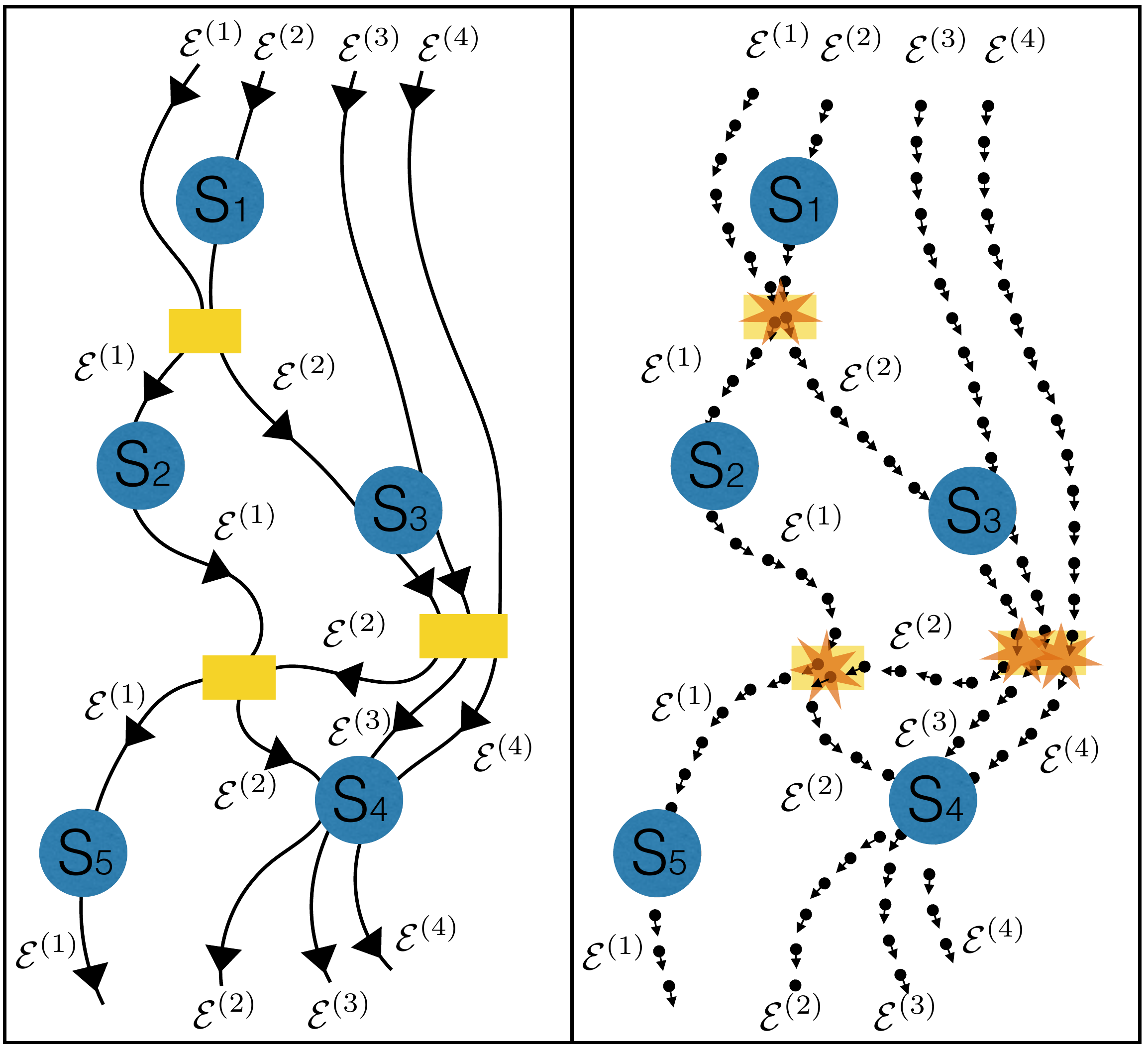}
\caption{Left panel: Pictorial representation of the typical QCS model we are considering here:   a collection ${\cal S}$  of quantum subsystems  $S_1$, $S_2$, $\cdots$, $S_M$ (gray circles in the figure) interact unidirectionally by exchanging signals through an 
 oriented network of environmental channels  ${\cal E}^{(1)}$, ${\cal E}^{(2)}$, $\cdots$, ${\cal E}^{(K)}$  which may interfere when intercepting (gray/yellow elements). Right panel: collisional model description of the scheme: the propagation
 of signals along the network is represented in terms of sequence of ordered collisional events involving the quantum subsystem and a collection of quantum information carriers (black circles). Interference among the signals arises from collisions between carriers associated with different connecting paths.
\label{figure1}}
\end{figure}
%%%%%%%%%%

In the QCS models studied so far, the parties composing the systems  are typically assumed  to be organized to form an oriented  linear chain, each  acting as 
controller for the elements that follow along the line through the mediation of a single environmental channel. 
Here instead we shall consider more complex configurations where several subsystems interact, unidirectionally via 
 a network of mutually intercepting channels as shown in the left panel of Fig.~\ref{figure1}. In this scenario the QCS couplings while being intrinsically dissipative in nature, can be affected by interference effects which originate from the propagation of the controlling pulses along the network of connections (for instance in the case of figure, the signals from the subsystem $S_1$ split and recombine before reaching subsystem $S_5$). 
 Also, depending on the topology of the scheme, controlling signals from different parties (say the subsystems $S_2$ and $S_3$ of the figure) can merge before reaching a given idler ($S_4$). The study of such architectures is intriguing as it widens the possibility of 
 engineering system-bath coupling in quantum optical systems, which in turn may help in 
 dissipatively preparing quantum many-body states of matter~\cite{PUPILLO,pra78042307,prl106090502} with important consequences in the analysis of  non-equilibrium condensed matter physics problems~\cite{phasesindoqs,prl105015702,pra83013611,pra86033821} and  quantum information~\cite{pra65010101,prl851762,njp16045014,prl107080503,PhysRevA.89.022335}.
 Aim of the present work is to derive a mathematical framework that incorporate these phenomena in a consistent way. For this purpose we shall adopt the collisional approach to QCS  introduced in Refs.~\cite{PhysRevLett.108.040401,0953-4075-45-15-154003}. Accordingly each unidirectional channel  forming the network of connections  is described in terms of a collection of sub-environments (quantum carriers) that evolve in time  stroboscopically through a series of time-ordered  collisions involving
 the various subsystems  -- see right panel of Fig.~\ref{figure1}. Interference effects are also described  in terms of collisions, this time involving
 carriers associated with different channels (e.g. the red and black carriers of the figure). Similar cascade networks could also be studied in the Heisenberg picture within the so called  {\it input-output} formalism \cite{gardinerbook, gard-zoll-book}, from which in principle a master equation could be derived using quantum stochastic calculus \cite{gardinerbook, Vogel2006}. The collisional model presented in this work allows to directly obtain the desired master equation and, being based on a simple and operational model of dissipation, naturally generates a Markovian completely positive dynamics without the necessity of introducing further hypothesis and approximations typical of other microscopic derivations.

Here is the outline of the paper: in Sec. \ref{sec:model} we review briefly the collisional approach to QCS of Ref.~\cite{PhysRevLett.108.040401,0953-4075-45-15-154003}
and adapt it for writing  the master equation of our model. The resulting expression is then cast in standard Gorini-Kossakowski-Sudarshan-Lindblad (GKSL) form~\cite{KOS,LIN,GO,breuer-petruccione} in Sec.~\ref{DD}. 
Building from these results, in Sec. \ref{sec:applications} we describe the arising of interference effects in the model, by discussing some specific examples. In particular  in Sec. \ref{sec:interferometer} we deal with a Mach-Zender-like interferometer, showing how with a phase shift it is possible to modify the effective temperature felt by the second optical cavity. Then in Sec. \ref{sec:example2} we turn to a 
configuration of three cavities where we show how, by appropriately exploiting interference effects, it is possible to have a system with only first-neighbor interactions.
The paper then ends with Sec.\ref{SEC:CON} where we draw conclusions and give an outlook for future works, and with the Appendices where we present some technical
derivations.

\section{\label{sec:model}The model}

In the collisional model approach \cite{RevModPhys.86.1203,doi:10.1007/s11080-005-0488-0,PhysRevA.61.022301,PhysRevA.65.042105,0034-4885-77-9-094001,PhysRevA.87.040103,arXiv:1608.03497v1} to open quantum systems dynamics
the environment is represented as  a large  many-body quantum system
whose constituents  (quantum information carriers or carriers in the following)  interact with the system of interest via an ordered sequence of impulsive unitary transformations (collisional events).
This yields a, time-discrete,  stroboscopic  evolution which can then be turned into a continuous time dynamics
by properly sending to infinite the number of collisions and to zero the time interval among them while keeping constant their product. By means of collisional models it is possible to derive both Markovian \cite{PhysRevA.61.022301,PhysRevA.65.042105} and non-Markovian \cite{0034-4885-77-9-094001,PhysRevA.87.040103,arXiv:1608.03497v1} master equations.
In this paper we take our steps from the collisional approach to QCS presented in Refs.~\cite{PhysRevLett.108.040401,0953-4075-45-15-154003}, generalizing it to include network configurations similar to the one presented in the left panel of Fig.~\ref{figure1}.

  %%%%%%
  \begin{figure}[!t]
\centering
\includegraphics[scale=0.3]{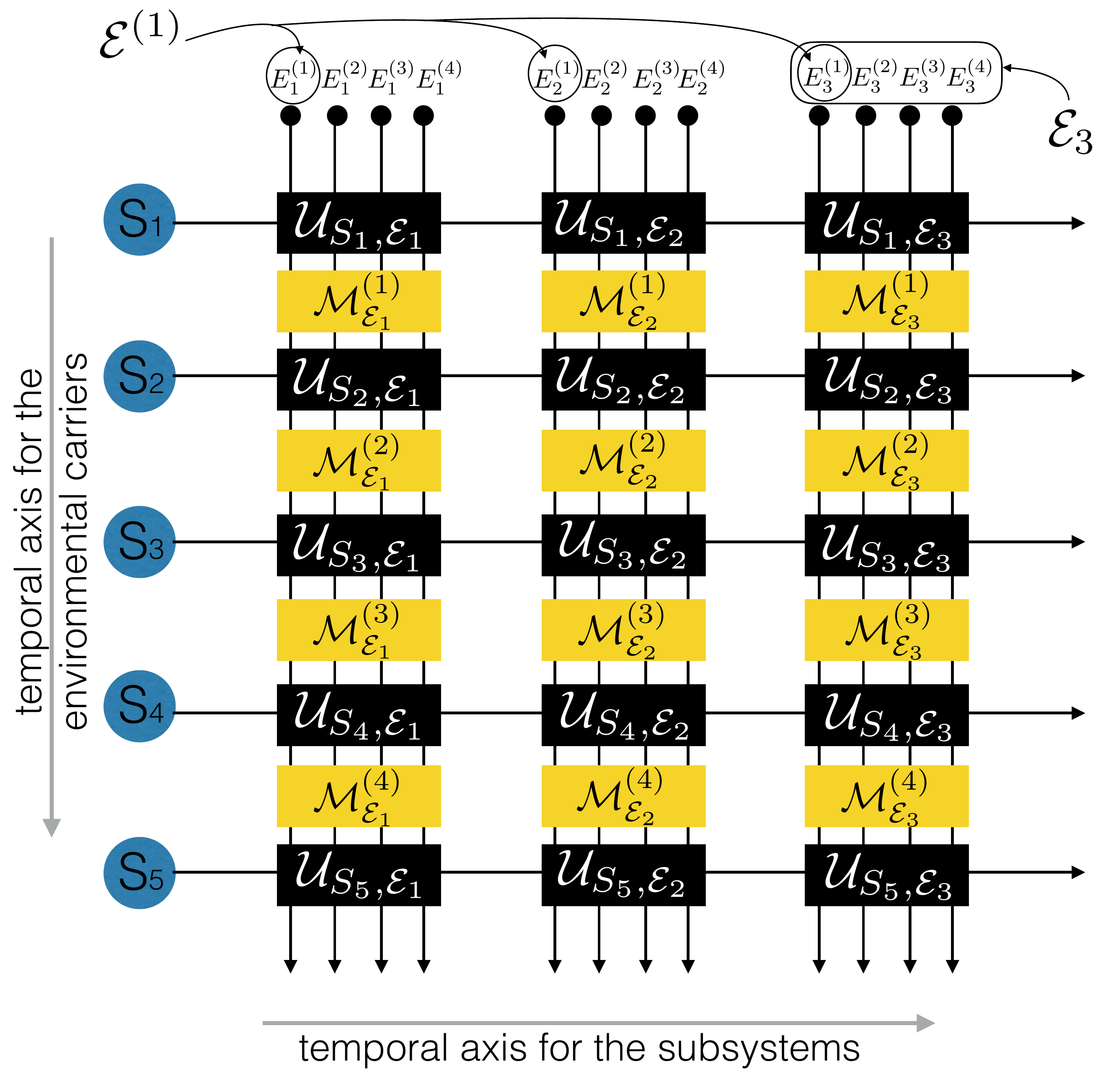}
\caption{Flowchart representation of the couplings in a  QCS network in the collisional model approach. The evolution of the
quantum carriers  $\{ E^{(k)}_n;k=1,2,\cdots,  n=1,2, \cdots \}$ representing the channels evolve in time from top to bottom, while the quantum subsystems 
$\{ S_{m}; m=1,2,\cdots\}$ evolve from left to right. The black elements represent collisional events between one of the subsystems and the carriers; the
yellow (pale gray) elements instead represent the dynamical evolution  of the carriers among two consecutive collisional events (possibly including interactions among carriers of different species). Notice that the upper index and lower index of carrier $E_{n}^{(k)}$  refer, respectively, to the environmental channel  (i.e. the channel ${\cal E}^{(k)}$ in this case) and the time group (i.e.  ${\cal E}_n$) it belongs~to.  \label{figure2}}
\end{figure}
%%%%%%%%%% 

 To this aim we consider a system ${\cal S}$ made out of $M$ (not necessarily identical)  subsystems $S_1$, $S_2$, $\cdots$, $S_M$ (e.g. $M$ optical cavities). Similarly to the scheme of Fig.~\ref{figure1}, they  are connected via a network of QCS  interactions  in such a way that
 for each $m=1,\cdots, M$, the element $S_m$ is  capable of controlling all the elements $S_{m'}$ with $m' >m$ without being affected by their dynamics, the coupling  being provided by a collection of unidirectional environmental channels   ${\cal E}^{(1)}$, ${\cal E}^{(2)}$, $\cdots$, ${\cal E}^{(K)}$   which intercept to form a graph.  
In what follows each of these channels are represented in terms of a long, ordered string of  quantum carriers which act as mediators of the interactions, propagating along the network and experiencing impulsive interactions (collisional events) 
 with the system elements as sketched on the right panel of Fig.~\ref{figure1}.
 Specifically, for $k=1,\cdots, K$, the $k$-th channel   ${\cal E}^{(k)}$
is  described by the carriers 
  $\{ E^{(k)}_n; n=1,2, \cdots \}$, the subscript $n$ indicating the order with which they start interacting with~${\cal S}$. Accordingly 
  we find it convenient to regroup these elements into sets which includes those that posses the same value of $n$ independently from the channel they belong to, e.g. the set ${\cal E}_1 :=   \{ E^{(1)}_1, 
  E^{(2)}_1, \cdots, E^{(K)}_1\}$, the set ${\cal E}_2 :=   \{ E^{(1)}_2, 
  E^{(2)}_2, \cdots, E^{(K)}_2\}$, and so on and so forth. 
This way, neglecting the time it takes from one carrier to move from one element of ${\cal S}$ to the next, we can use  the label $n$ as the discrete temporal coordinate of the model (more on this on the following paragraphs).
  In particular, 
indicating with $\hat{U}_{S_m,{\cal E}_n}$  the unitary operator associated with the collisional event that couples  $S_{m}$ and the carriers which enters at the $n$-th temporal step, i.e. the carriers of  ${\cal E}_n$,
   the causal structure of the model is enforced by imposing that such operator should precede  $\hat{U}_{S_{m+1},{\cal E}_n}$
  (meaning that $S_{m+1}$ sees  ${\cal E}_n$ only after it has interacted with $S_m$) 
  and $\hat{U}_{S_{m},{\cal E}_{n+1}}$ (meaning that the element of ${\cal E}_n$ enters the network before those of ${\cal E}_{n+1}$) -- 
  the relative ordering of $\hat{U}_{S_{m+1},{\cal E}_n}$  and  $\hat{U}_{S_{m},{\cal E}_{n+1}}$ being instead irrelevant as they act on different
  systems and hence commute.  
The unitaries   $\hat{U}_{S_m,{\cal E}_n}$'s  trigger the dissipative evolution of ${\cal S}$ which is responsible for the QCS dynamics. In our model they are interweaved with Completely
Positive and Trace preserving (CPT) super-operators~\cite{HolevoBOOK} acting on the quantum carriers only, which describe the propagation of signals along the channels and (possibly) their mutual interactions.  In particular, in what follows we shall use the symbol ${\cal M}^{(m)}_{{\cal E}_n}$ to indicate  the CPT map which
acts on the carries of the set ${\cal E}_n$ 
  after the collisional event that couples them with $S_{m}$ and before the one that instead couples them with  $S_{m+1}$  -- see Fig.~\ref{figure2}.
 A convenient way to express the resulting evolution  
is obtained by introducing the density matrix $\hat{R}(n)$ which describes the joint state of ${\cal S}$ and of the first $n$-th carriers of all
channels (i.e. the carriers belonging to the sets ${\cal E}_1$, ${\cal E}_2$, $\cdots$, ${\cal E}_n$) after they have interacted. From the above construction, the relation between
 such state and its evolved counterpart  $\hat{R}(n+1)$ can then be expressed as 
\begin{equation}
\label{eq:recursive_density_matrix_equation}
\hat{R}(n+1)=\mathcal{C}_{\mathcal{S},{\cal E}_{n+1}}(\hat{R}(n)\otimes \hat{\eta}_{{\cal E}_{n+1}})\;,
\end{equation}
where $\hat{\eta}_{{\cal E}_{n+1}}$ indicates the input state of the elements of ${\cal E}_{n+1}$ when they enter the network,
while $\mathcal{C}_{\mathcal{S},{\cal E}_{n+1}}$ is the super-operator associated with the collisional events they participate.  Explicitly, 
using the short hand notation \begin{eqnarray}  
\label{SHORT} 
\overleftarrow{\Pi}_{m=1}^M {\cal A}^{(m)} =  {\cal A}^{(M)}  {\cal A}^{(M-1)} \cdots  {\cal A}^{(2)} {\cal A}^{(1)}\end{eqnarray}  to represent
 the ordered product of the symbols $\{ {\cal A}^{(1)}, {\cal A}^{(2)}, \cdots, {\cal A}^{(M)}\}$
 this is given by 
\begin{eqnarray} \label{DEFC}&& \mathcal{C}_{\mathcal{S},{\cal E}_{n}}=
\overleftarrow{\Pi}_{m=1}^M [\mathcal{M}^{(m)}_{{\cal E}_{n}} \circ \mathcal{U}_{S_m,{\cal E}_{n}}] \;, 
\end{eqnarray}
where ``$\circ$" indicates composition of super-operators  and where for each $m=1,\cdots, M$ 
 the symbol $\mathcal{U}_{S_m,{\cal E}_{n}}$ 
indicates  the super-operator counterpart of the unitary transformation $\hat{U}_{S_m,{\cal E}_{n}}$, i.e. 
\begin{eqnarray} \mathcal{U}_{S_m,{\cal E}_{n}}(\cdots) =
\hat{U}_{S_m,{\cal E}_{n}} ( \cdots) \hat{U}_{S_m,{\cal E}_{n}}^\dag\label{SUPU}\;. \end{eqnarray}
Few remarks are mandatory at this point: 
\begin{itemize} 
\item[i)]  the density matrices $\hat{R}(n)$ and $\hat{R}(n+1)$ operate on different spaces (indeed $\hat{R}(n+1)$  applies also to the carriers of the set ${\cal E}_{n+1}$ while
$\hat{R}(n)$ does not). What is relevant for us is the fact that by taking the partial trace over the carriers they give us the temporal evolution of the system
of interest at the various step of the process. In particular \begin{eqnarray} \hat{\rho}(n) := \Big\langle  \hat{R}(n)\Big\rangle_{\cal E} =  \mbox{Tr}_{\cal E}[ \hat{R}(n)] \label{DEFRHO}\;, \end{eqnarray} is the joint state of the subsystems ${\cal S}$ at the $n$-th time step;
\item[ii)] as already mentioned in our analysis the time it takes for a carrier to move from one collision to the next is assumed to be negligible,  only the causal ordering of these events being preserved. Accordingly in Fig.~\ref{figure2} time flows from left to right for all the $S_j$  synchronously. This assumption is introduced because, differently to the case of a simple linear chain of cascaded systems~\cite{gard-zoll-book}, when dealing with multiple channels one cannot eliminate the delay time by simply shifting the time origin of each subsystem. Actually, significant delay times can give rise to non-Markovian effects~\cite{PhysRevA.87.040103}, whose study goes beyond the goal of the present work.%only not to burden the notation and  can be easily circumvent 
%by including effective delay times in the definition of the density matrix $\hat{\rho}(n)$ (see e.g. ....);
\item[iii)] in writing Eq.~(\ref{eq:recursive_density_matrix_equation}) we are implicitly assuming that
the input state of the carriers factorizes with respect to the grouping ${\cal E}_1$, ${\cal E}_2$, $\cdots$, i.e. no correlations are admitted among 
carriers which enters the scheme at different time steps. Yet, at this level, the model still admits the possibility of correlations among carriers of different channels. 
In what follows we shall however enforce a further constraint that limits the choices of  the input $\hat{\eta}_{{\cal E}_{n}}$, see next point and  Eq.~(\ref{eq:stability_condition}) below. 
\item[iv)] In the original QCS model of Fig.~\ref{figure1} the unidirectional channels ${\cal E}^{(1)}$, ${\cal E}^{(2)}$ and ${\cal E}^{(K)}$ form a stationary medium  which contributes to the dynamics only by allowing  signals from
one subsystem to propagate to the next one (in other words in the absence of the interactions with the elements of ${\cal S}$ they will not present any temporal evolution).  To enforce this special  character in the collisional model we  require it to be translationally invariant with respect to the index $n$, e.g. imposing  that all the input states 
$\hat{\eta}_{{\cal E}_{1}}$, $\hat{\eta}_{{\cal E}_{2}}$, $\cdots$,  $\hat{\eta}_{{\cal E}_{n}}$ of the carriers sets  
${\cal E}_{1}$, ${\cal E}_{2}$, $\cdots$, ${\cal E}_n$ coincide,  and that for given $m$  the unitary couplings 
 $\mathcal{U}_{S_m,{\cal E}_{n}}$ and  the maps $\mathcal{M}^{(m)}_{{\cal E}_{n}}$ should be independent from $n$.
This hypothesis can however be relaxed~\cite{0953-4075-45-15-154003} with the condition that the change in the coupling is slow compared to the characteristic time scale of the systems $S_m$.
\end{itemize}

\subsection{The continuous time limit} \label{subsectionA} 

By solving the recursive equation~(\ref{eq:recursive_density_matrix_equation})  and taking the partial trace as in Eq.~(\ref{DEFRHO}) one obtains
a collection of  density matrices  $\hat{\rho}(0)$, $\hat{\rho}(1)$, $\cdots$, $\hat{\rho}(n)$, which provides 
 an effective description of the temporal evolution of the joint state of the subsystems ${S}_1$,  $S_2$, $\cdots$,
 $S_M$ in the presence of a collection of quantum carriers that connects them  through a network of unidirectional channels. 
Such stroboscopic representation of the dynamics can be turned into a continuous time description by taking a proper limit in which the number of collisions per second
experienced by the element of ${\cal S}$  goes to infinity~\cite{PhysRevLett.108.040401,0953-4075-45-15-154003}. Accordingly  we write the interaction unitaries as
 \begin{eqnarray} \label{UNITARY} 
  \hat{U}_{S_m,{\cal E}_{n}} =\exp[-i g \sum_{k=1}^K \hat{H}_{S_m,{ E}_n^{(k)}} \Delta t]\;, 
  \end{eqnarray}  
where  $g$ is a coupling constant that we shall use to gauge the intensity of the system-carrier interactions, $\Delta t$ is  the duration of a single  collisional event, and 
where 
\begin{eqnarray}  \label{COUPH} 
\hat{H}_{S_m,{ E}_n^{(k)}}&=&\sum_{\ell}\hat{A}^{(\ell,k)}_{S_m}\otimes \hat{B}^{(\ell,m)}_{E_{n}^{(k)}}\;, 
\end{eqnarray}
is the most general Hamiltonian describing the interactions between  $S_m$ and ${E}_n^{(k)}$, with  $\hat{A}^{(\ell,k)}_{S_m}$ and  $\hat{B}^{(\ell,m)}_{E_{n}^{(k)}}$ nonzero operators acting locally on such systems 
respectively~\cite{breuer-petruccione}. 
 Next we 
 take the product $g \Delta t$ to be a small quantity and expand our equations up to the second order in such term. In this regime, upon tracing  upon the degree of freedom 
 of the carriers, Eq.~(\ref{eq:recursive_density_matrix_equation}) yields the identity 
  \begin{eqnarray}
&&\frac{\hat{\rho}(n+1)-\hat{\rho}(n)}{\Delta t} =  -ig\sum_{m,k,\ell}
\gamma_{m (k)}^{(\ell)}   \Big[\hat{A}_{S_m}^{(\ell,k)},\hat{\rho}(n)\Big]_{-}
   \nonumber\\
&&\; +g^2\Delta t \; \left\{ \sum_{m=1}^M{\cal L}_{m}(\hat{\rho}(n))  
+\sum_{m'=m+1}^{M}\sum_{m=1}^{M-1}{\cal D}_{m,m'}(
\hat{\rho}(n)) \right\}  \nonumber 
\\ 
&& \qquad +\mathcal{O}(g^3\Delta t^2), \label{RHOAPPR} 
\end{eqnarray}
with  \begin{eqnarray}
{\cal L}_m(\cdots)  &=& 
\frac{1}{2} \sum_{k,k'=1}^K\sum_{\ell,\ell'}   \gamma_{m(kk')}^{(\ell,\ell')} 
 \;\Big\{  2  \hat{A}_{S_m}^{(\ell,k)} (\cdots) \hat{A}_{S_m}^{(\ell',k')}
\nonumber \\ \label{DEFLLOC} 
&&\qquad -\Big[ \hat{A}_{S_m}^{(\ell',k')} \hat{A}_{S_m}^{(\ell,k)}  ,\cdots  \Big]_{+} \Big\},\end{eqnarray} 
and for $m'>m$, 
\begin{eqnarray}
{\cal D}_{m\rightarrow m'}(\cdots)& =& 
\sum_{k,k'=1}^K \sum_{\ell,\ell'}
\Big\{
\zeta_{mm'(kk')}^{(\ell,\ell')} 
\; \hat{A}_{S_m}^{(\ell,k)}\Big[\cdots,\hat{A}_{S_{m'}}^{(\ell',k')}\Big]_{-} \nonumber 
\\ &&  - \; 
\xi_{mm'(kk')}^{(\ell,\ell')} \; 
\Big[\cdots, {\hat{A}_{S_{m'}}^{(\ell',k')}} \Big]_{-}{\hat{A}_{S_m}^{(\ell,k)}} \Big\}, 
\label{DEFDLOC}
\end{eqnarray}
where $\Big[ \cdots, \cdots \Big]_{\pm}$ represent the commutator ($-$) and anti-commutator ($+$)  brackets respectively.
In the above expressions  $\gamma_{m (k)}^{(\ell)}$, $\gamma_{m(kk')}^{(\ell,\ell')}$, $\zeta_{mm'(kk')}^{(\ell,\ell')}$, and $\xi_{mm'(kk')}^{(\ell,\ell')}$ are complex coefficients which depend upon correlation term of the 
  input state of the carriers  (see Appendix~\ref{sec:derivation} for the explicit definitions). 

The continuous time limit is finally obtained sending to infinity $n$ of collisions while the time interval $\Delta t$ of each collision goes to zero and the coupling constant $g$ explodes in such a way that:
\begin{equation} \label{LIMIT} 
\lim_{\Delta t\rightarrow0^+}n\Delta t=t\;,\qquad \lim_{\Delta t\rightarrow0^+}g^2\Delta t=\gamma\;,
\end{equation}
with $\gamma$ being a positive constant which set the time scale of the model. 
Notice that the last assumption could lead to problem in the first-order term of the series expansion of Eq.~(\ref{RHOAPPR}), whose contribution to the final expression would explode. Such instability is a typical trait in the derivation of master equations~\cite{breuer-petruccione} for open quantum systems. It
can be solved by imposing a  stability condition~\cite{PhysRevLett.108.040401,0953-4075-45-15-154003}  for the environmental  degree of freedom  of the system, i.e.  by requiring that 
the input carrier states $\hat{\eta}_{{\cal E}_n}$ and their evolved counterparts along the network should not be influenced (at first order)  by the  collisions with the subsystems. 
This is consistent with the description of the environmental channels as composed by many small sub-environments all in the same reference state that interacts weakly with the subsystems. 
In the standard derivation of master equations  such stability condition is usually assumed as well, and it amounts to the possibility of approximating the joint density matrix as a tensor product between the reduced density matrix of the system and the one of the environment at any time. In our case this corresponds to nullify the 
coefficients $\gamma_{m (k)}^{(\ell)}$ appearing in the rhs of  Eq.~(\ref{RHOAPPR}), i.e. by imposing (see Eq.~(\ref{LAMBDA1}) of the Appendix~\ref{sec:derivation}) 
\begin{eqnarray}
&&\Big\langle   \hat{B}^{(\ell,1)}_{E^{(k)}_{n}} \; \hat{\eta}_{{\cal E}_n} \Big\rangle_{{\cal E}} =0 \;, \nonumber \\ 
&&\Big\langle   \hat{B}^{(\ell,2)}_{E^{(k)}_{n}} \; 
\mathcal{M}^{(1)}_{{\cal E}_n}(\hat{\eta}_{{\cal E}_n}) \Big\rangle_{{\cal E}} =0 \;, \nonumber \\ 
&&\Big\langle   \hat{B}^{(\ell,3)}_{E^{(k)}_{n}} \; [\mathcal{M}^{(2)}_{{\cal E}_n} \circ 
\mathcal{M}^{(1)}_{{\cal E}_n}] (\hat{\eta}_{{\cal E}_n}) \Big\rangle_{{\cal E}} =0 \;, \nonumber \\  && \qquad \vdots \nonumber \\ 
&& \Big\langle   \hat{B}^{(\ell,m)}_{E^{(k)}_{n}}  [ 
\mathcal{M}^{(m-1)}_{{\cal E}_n}\circ \cdots \circ \mathcal{M}^{(1)}_{{\cal E}_n}](\hat{\eta}_{{\cal E}_n}) \Big\rangle_{{\cal E}} = 0 \;, \label{eq:stability_condition}
\end{eqnarray}
for all $k=1,\cdots, K$,  for all $m=1,\cdots, M$, and for all $\ell$ ($\hat{B}^{(\ell,m)}_{E^{(k)}_{n}}$ being the carriers operators which participate to the coupling Hamiltonian~(\ref{COUPH})). 
% and where for $m>1$, $[ \overleftarrow{\Pi}_{m'=1}^{m-1} 
%\mathcal{M}^{(m')}_{{\cal E}_n}](\hat{\eta}_{{\cal E}_n})$ represents the evolution of the carriers states along the network in the absence of interactions with the element of ${\cal S}$ (for $m=0$ 

By enforcing the condition (\ref{eq:stability_condition}), Eq.~(\ref{RHOAPPR}) finally can be casted in the following
differential form 
\begin{eqnarray}
\label{eq:cascade_master_equation}
\frac{\partial\hat{\rho}(t) }{\partial t}&=&\gamma \; {\cal C}(\hat{\rho}(t)) \;, \end{eqnarray}
with ${\cal C}$ the QCS super-operator 
\begin{eqnarray} 
{\cal C}(\cdots) = \sum_{m=1}^M\mathcal{L}_{m}(\cdots)+ \sum_{m'=m+1}^{M}\sum_{m=1}^{M-1} \mathcal{D}_{m\rightarrow m'}(\cdots)\;. \nonumber \\
\label{DEFC} 
\end{eqnarray}
Equation~(\ref{eq:cascade_master_equation})  is a Markovian  master equation which describes the dynamical evolution of the joint density matrix $\hat{\rho}(t)$ for the system of interest ${\cal S}$.
The term on the rhs is the generator of the dynamics and can be casted in GKSL form~\cite{KOS,LIN,GO,breuer-petruccione} by properly reorganizing the various contributions (see next section). It is however worth analyzing the causal structure of the model a bit further by looking directly  at the expression presented in~\eqref{DEFC}. 
On the one hand we have the terms $\mathcal{L}_m$ which describe local effects of the interaction between the various element of ${\cal S}$ and the environment: they are  not capable of creating correlations among the ${\cal S}_m$'s and only account for dissipative behaviors.  On the other hand  the non-local terms $\mathcal{D}_{m\rightarrow m'}$ describe the interaction between the $m$-th and $m'$-th subsystem (with $m'>m$) originating by the propagation of the carries from the former to the latter. In principle these are capable of building up 
correlations among the various elements of ${\cal S}$. However, at variance with what would happen with a direct Hamiltonian interaction, such couplings are intrinsically 
asymmetric in agreement with the cascade structure of the network of connections. In particular one may observe that by tracing over ${S}_{m'}$ the term $\mathcal{D}_{m\rightarrow m'}(\hat{\rho}(t))$ always
nullifies, i.e. \begin{eqnarray}
\mbox{Tr}_{S_{m'}} [ \mathcal{D}_{m\rightarrow m'}(\hat{\rho}(t))] =0\;,\end{eqnarray}  while this 
 is not necessarily the case when the same term is traced over $S_m$.   This implies, for instance, that the reduced density matrix $\hat{\rho}_1(t)$ of the first element of ${\cal S}$ (the one which in principle controls all the others without being controlled by them) evolves in time without being affected by the presence of the latter. Similarly
 the evolution of the first $m$ elements of ${\cal S}$ does not depend upon the remaining ones. 

The derivation of Eq.~(\ref{eq:cascade_master_equation}) we have presented here closely follows the one of Ref.~\cite{PhysRevLett.108.040401}. The main difference with the latter is the inner structure of the generators ${\cal L}_m$ and ${\cal D}_{m,m'}$ which in our case includes contributions from multiple unidirectional channels as indicated by the sum over the indexes $k$ and $k'$ of Eqs.~(\ref{DEFLLOC}) and (\ref{DEFDLOC}). As it will be clear discussing some explicit examples (see next Section) this is what allows us to 
account for interference effects that originate with the signals propagation through the network. 

\subsection{Standard GKSL form and effective Hamiltonian couplings} \label{DD}

The decomposition  of the coupling Hamiltonians presented in Eq.~(\ref{COUPH}) is clearly not unique.
Alternatives  can be obtained  by replacing the $\hat{A}^{(\ell,k)}_{S_m}$'s (resp. the $\hat{B}^{(\ell,m)}_{E_{n}^{(k)}}$'s) with proper linear combinations of the same objects for instance by expanding them into an operator basis.
The  master equation~(\ref{eq:stability_condition}) clearly does not depend on this choice as it derives from a pertubative expansion on the coupling parameter $g$ which enters in the model as a multiplicative factor of $\hat{H}_{S_m,{ E}_n^{(k)}}$, and  from the stability conditions~(\ref{eq:stability_condition}), which are explicitly invariant under linear combinations of the $\hat{B}^{(\ell,m)}_{E_{n}^{(k)}}$'s. In this section we shall invoke this freedom assuming the $\hat{A}^{(\ell,k)}_{S_m}$'s and the $\hat{B}^{(\ell,m)}_{E_{n}^{(k)}}$'s
to be self-adjoint (a possibility which is allowed by the fact that $\hat{H}_{S_m,{ E}_n^{(k)}}$ has to be self-adjoint as well).  This working hypothesis is not fundamental but, as
pointed out in Refs.~\cite{PhysRevLett.108.040401,0953-4075-45-15-154003}, turns out to be useful as it 
makes explicit some structural properties of the resulting super-operators, 
ensuring for instance the identities
\begin{eqnarray}
\gamma_{m(kk')}^{(\ell,\ell')}&=&\left[\gamma_{m(k'k)}^{(\ell',\ell)}\right]^*\;,    \label{fffd}  \\ 
\xi_{mm'(kk')}^{(\ell,\ell')}&=&\left[\zeta_{mm'(kk')}^{(\ell,\ell')}\right]^*\;,  \label{fffd1} 
\end{eqnarray} 
as evident from  Eqs.~(\ref{eq:general_local_coefficient})-(\ref{eq:general_interaction_coefficient1}) of the Appendix.
Our aim is to exploit these properties  to generalize the analysis  of Ref.~\cite{1367-2630-14-6-063014} by casting the QCS super-operator~(\ref{DEFC})
into 
an explicit  standard  GKSL form~\cite{breuer-petruccione}, i.e. as the sum of an  effective Hamiltonian term plus a collection of purely dissipative contributions
\begin{eqnarray} 
\label{eq:cascade_master_equationSTANDARD}
{\cal C}(\cdots) &=& - i [\hat{H},(\cdots) ] 
\\
&+& \sum_{i}  2  \hat{L}^{(i)} (\cdots) \hat{L}^{(i)\dag} -  
\Big[ \hat{L}^{(i)\dag}  \hat{L}^{(i)} , (\cdots)  \Big]_{+} \;,  \nonumber 
\end{eqnarray} 
with $\hat{H}$ being self-adjoint and with the $\hat{L}^{(i)}$'s being a collection of operators acting on ${\cal S}$.
In Ref.~\cite{1367-2630-14-6-063014} this trick was used to show that 
a collection of two-level atoms
 coupled in QCS fashion via an unidirectional optical fiber, initialized at zero temperature, can be  described as  originating from an effective two-body coupling Hamiltonian 
with chiral symmetry.

We start by focusing on the local contributions of  Eq.~(\ref{eq:cascade_master_equation}). Indicating with $j$ the joint index $(\ell,k)$, Eq.~(\ref{fffd})  implies that, for each $m$ assigned, the matrix $\Theta_{jj'}$  of  elements $\gamma_{m(kk')}^{(\ell,\ell')}/2$ is Hermitian, i.e. $\Theta_{jj'} = \Theta_{j'j}^*$. Furthermore, by direct inspection of Eq.~(\ref{eq:general_local_coefficient}) one can easily prove that, being $\hat{B}^{(\ell,m)}_{E_{n}^{(k)}}$ self-adjoint,  such matrix is also semi-positive definite. 
Accordingly Eq.~(\ref{DEFLLOC}) can be expressed as a purely dissipative term 
%casted in a standard Lindblad form~\cite{breuer-petruccione}, i.e. 
 \begin{eqnarray}
{\cal L}_m(\cdots)  = 
\sum_{s}   \lambda_s  
 \;\Big\{  2  \hat{\Lambda}_{S_m}^{(s)}  (\cdots) \hat{\Lambda}_{S_m}^{(s)\dag} -\Big[ \hat{\Lambda}_{S_m}^{(s)\dag}  \hat{\Lambda}_{S_m}^{(s)}  ,\cdots  \Big]_{+} \Big\}, 
\nonumber \\ \label{normalL} 
\end{eqnarray} 
where  $\{ \lambda_s\}_s$ are  the  eigenvalues of $\Theta_{j,j'}$ and where 
we have introduced the
operators  
\begin{eqnarray} 
\hat{\Lambda}_{S_m}^{(s)} = \sum_{k,\ell} \; {v}_{(\ell,k),s} \;   \hat{A}_{S_m}^{(\ell,k)}\;, \end{eqnarray} 
with  ${v}_{j,s}$ being the unitary matrix which allows us to diagonalize  $\Theta_{j,j'}$, i.e. $\Theta_{j,j'} = \sum_{s} {v}_{j,s} \lambda_s {v}_{s,j'}^*$.  In the absence of the coupling
contributions ${\cal D}_{m,m'}$,  Eq.~(\ref{eq:cascade_master_equation}) will hence reduce to the standard form~(\ref{eq:cascade_master_equationSTANDARD})
with $\hat{H}=0$ and with the dissipative operators $\hat{L}^{(i)}$  being identified with $\sqrt{\lambda_s} \hat{\Lambda}_{S_m}^{(s)}$. 

Consider next the non-local contributions  of Eq.~(\ref{eq:cascade_master_equation}). 
 Due to their peculiar structure they cannot directly produce terms 
as those on the right hand side of  Eq.~(\ref{eq:cascade_master_equationSTANDARD}). 
We notice however  that for all $m' >m$  one can write 
\begin{eqnarray}
\hat{A}_{S_m}^{(\ell,k)}\Big[\cdots,\hat{A}_{S_{m'}}^{(\ell',k')}\Big]_{-} \nonumber = 
- \frac{1}{2} \Big[\hat{A}_{S_m}^{(\ell,k)} \hat{A}_{S_{m'}}^{(\ell',k')} ,\cdots \Big]_{-} \\
+ \hat{A}_{S_m}^{(\ell,k)} \cdots \hat{A}_{S_{m'}}^{(\ell',k')} - \frac{1}{2} \Big[\hat{A}_{S_m}^{(\ell,k)} \hat{A}_{S_{m'}}^{(\ell',k')} ,\cdots \Big]_{+}  \;,\nonumber 
\end{eqnarray}
and 
\begin{eqnarray} 
\Big[\cdots,\hat{A}_{S_{m'}}^{(\ell',k')}\Big]_{-}\; \hat{A}_{S_m}^{(\ell,k)} \nonumber = 
- \frac{1}{2} \Big[ \hat{A}_{S_{m'}}^{(\ell',k')} \hat{A}_{S_m}^{(\ell,k)} ,\cdots \Big]_{-} \\
-  \hat{A}_{S_{m'}}^{(\ell',k')} \cdots \hat{A}_{S_m}^{(\ell,k)}   + \frac{1}{2} \Big[\hat{A}_{S_{m'}}^{(\ell',k')} \hat{A}_{S_m}^{(\ell,k)}  ,\cdots \Big]_{+} \;, 
\end{eqnarray}
 which simply follow from the fact that $\hat{A}_{S_m}^{(\ell,k)}$, $\hat{A}_{S_{m'}}^{(\ell',k')}$ operate on different quantum systems and hence commute. 
Replacing these identities into Eq.~(\ref{DEFDLOC}) we can write  
\begin{eqnarray}
{\cal D}_{m\rightarrow m'}(\cdots)& =&  - i \Big[\hat{H}_{m,m'}, (\cdots)\Big]_- + \Delta {\cal L}_{m,m'}(\cdots)  \;.\nonumber \\ \label{DEFDLOCnew}
\end{eqnarray} 
In this expression the first contribution is an effective Hamiltonian term with 
\begin{eqnarray}
\hat{H}_{m,m'} &=&  \sum_{k,k'=1}^K \sum_{\ell,\ell'}
\frac{\xi_{mm'(kk')}^{(\ell,\ell')}-\zeta_{mm'(kk')}^{(\ell,\ell')}}{2 i } 
\; \hat{A}_{S_m}^{(\ell,k)} \otimes \hat{A}_{S_{m'}}^{(\ell',k')} \nonumber \\
&=& \sum_{k,k'=1}^K \sum_{\ell,\ell'}
\mbox{Im}[ \xi_{mm'(kk')}^{(\ell,\ell')}] 
\; \hat{A}_{S_m}^{(\ell,k)} \otimes \hat{A}_{S_{m'}}^{(\ell',k')}  \;, \label{CHIRHAM} 
\end{eqnarray}
where in the second line we used~(\ref{fffd}). The second contribution on the right hand side of~(\ref{DEFDLOCnew}) instead 
features the super-operator
\begin{eqnarray} 
&& \Delta{\cal L}_{m,m'}(\cdots) = \sum_{m_1,m_2=m,m'}  \sum_{k_1,k_2=1}^K \sum_{\ell_1,\ell_2}  \Delta D_{m_1,m_2(k_1k_2)}^{(\ell_1,\ell_2)}\nonumber \\
&& \times \left\{ 2 \hat{A}_{S_{m_1}}^{(\ell_1,k_1)}  (\cdots) \hat{A}_{S_{m_2}}^{(\ell_2,k_2)}  -  \left[ \hat{A}_{S_{m_1}}^{(\ell_1,k_1)} \hat{A}_{S_{m_2}}^{(\ell_2,k_2)}  ,  (\cdots) \right]_+  \right\}  \nonumber \\
 \label{CHIRLA}
\end{eqnarray}
with coefficients
\begin{eqnarray} 
\Delta D_{m_1,m_2(k_1k_2)}^{(\ell_1,\ell_2)} =\frac{1}{2}  \left\{ 
\begin{array}{ll} 
\zeta_{m_1m_2(k_1k_2)}^{(\ell_1,\ell_2)} & \mbox{for $m_1 < m_2$} \\
0  & \mbox{for $m_1 = m_2$} \\
\xi_{m_2m_1(k_2k_1)}^{(\ell_2,\ell_1)} & \mbox{for $m_1  > m_2$.}
\end{array} \right. \label{coef111} 
\end{eqnarray} 
One may notice that  indicating with $j$ the joint index $(\ell,k,m)$, then from Eq.~(\ref{fffd}) it follows that the matrix $\Delta\Omega_{j,j'}$ of elements
$\Delta D_{m_1,m_2(k_1k_2)}^{(\ell_1,\ell_2)}$ is Hermitian, i.e. $\Delta D_{m_1,m_2(k_1k_2)}^{(\ell_1,\ell_2)}= \left[ \Delta D_{m_2,m_1(k_2k_1)}^{(\ell_2,\ell_1)}\right]^*$.
Yet there is no guarantee that  $\Delta \Omega_{j,j'}$ is semi-positive definite  (an explicit counter-example will be presented in the next section) thus preventing one from directly expressing~(\ref{CHIRLA}) as a sum of dissipative contributions by diagonalization
of $\Delta \Omega_{j,j'}$ as we did for the local terms of~${\cal C}$. However by replacing Eq.~(\ref{DEFDLOCnew}) into~(\ref{DEFC}) we arrive to  
\begin{eqnarray} 
\label{eq:cascade_master_equationSTANDARDnew}
&&{\cal C}(\cdots) =  - i [\hat{H},(\cdots) ] 
+  \sum_{m_1,m_2=1}^M  \sum_{k_1,k_2=1}^K \sum_{\ell_1,\ell_2}  D_{m_1,m_2(k_1k_2)}^{(\ell_1,\ell_2)}\nonumber \\
&& \times \left\{ 2 \hat{A}_{S_{m_1}}^{(\ell_1,k_1)}  (\cdots) \hat{A}_{S_{m_2}}^{(\ell_2,k_2)}  -  \left[ \hat{A}_{S_{m_1}}^{(\ell_1,k_1)} \hat{A}_{S_{m_2}}^{(\ell_2,k_2)}  ,  (\cdots) \right]_+  \right\}  \nonumber \\
\end{eqnarray} 
where now $\hat{H}$ is the effective Hamiltonian 
\begin{eqnarray} \hat{H} = \sum_{m'=m+1}^M\sum_{m=1}^M  \hat{H}_{m,m'} \;,  \label{EFFEHAMI} 
\end{eqnarray} 
and where the coefficients $D_{m_1,m_2(k_1k_2)}^{(\ell_1,\ell_2)}$ are obtained from those of Eq.~(\ref{coef111}) by using  the elements $\gamma_{m(kk)}^{(\ell,\ell')}$ to fill the zero's on the $m_1$, $m_2$ diagonal, i.e. 
\begin{eqnarray} \label{COEFMEW} 
 D_{m_1,m_2(k_1k_2)}^{(\ell_1,\ell_2)} = \left\{ 
\begin{array}{ll} 
\Delta D_{m_1,m_2(k_1k_2)}^{(\ell_1,\ell_2)}  & \mbox{for $m_1 \neq  m_2$} \\ \\ 
\gamma_{m_1 (k_1,k_2)}^{(\ell_1,\ell_2)}/2 & \mbox{for $m_1 = m_2$.}
\end{array} \right.
\end{eqnarray} 
To complete the derivation of Eq.~(\ref{eq:cascade_master_equationSTANDARD}) 
one should prove  the  non-negativity of the matrix $\Omega_{j,j'} =D_{m_1,m_2(k_1k_2)}^{(\ell_1,\ell_2)}$  ($j$ being once more the joint index $(\ell,k,m)$). 
This is shown explicitly in App.~\ref{APPENDIXb}. Indicating hence with $\kappa_i (\geq 0)$ the eigenvalues of $\Omega_{j,j'}$ and with $w_{j,i}$ the elements of the unitary matrix that diagonalizes it (i.e. $\Omega_{j,j'} = \sum_{s} w_{j,i} \kappa_i w_{i,j'}^*$) 
we can finally identify the operators $\hat{L}^{(i)}$ of Eq.~(\ref{eq:cascade_master_equationSTANDARD}) with
\begin{eqnarray}\label{LLLLLLL}
\hat{L}^{(i)} = \sqrt{\kappa_i} \; \sum_{\ell,k,m} w_{(\ell,k,m),i} \; \hat{A}_{S_{m}}^{(\ell,k)} \;. 
\end{eqnarray} 

A final remark before concluding the section: 
as already mentioned in deriving the above results we find it convenient to assume the operators $\hat{A}^{(\ell,k)}_{S_m}$ and  $\hat{B}^{(\ell,m)}_{E_{n}^{(k)}}$ to be self-adjoint.
Yet the analysis presented here is still valid even when this assumption does not hold 
- simply some of the structural properties of the involved mathematical objects 
are less explicit. In particular the eigenvalues of the matrices~(\ref{coef111}) and (\ref{COEFMEW}) can be shown to be independent from the decomposition adopted
in writing~(\ref{COUPH}) (the associated matrices being related by similarity transformations). 

\section{\label{sec:applications}Interference Effects} 
Here we present a couple of  examples of QCSs which enlighten the arising of interference effects during the propagation of signals  on a network of unidirectional connections
and how they can be used to externally tune the couplings among the various subsystems. 
\subsection{\label{sec:interferometer}Example 1: Mach-Zehnder model}
\begin{figure}[!t]
\centering
\includegraphics[scale=0.27]{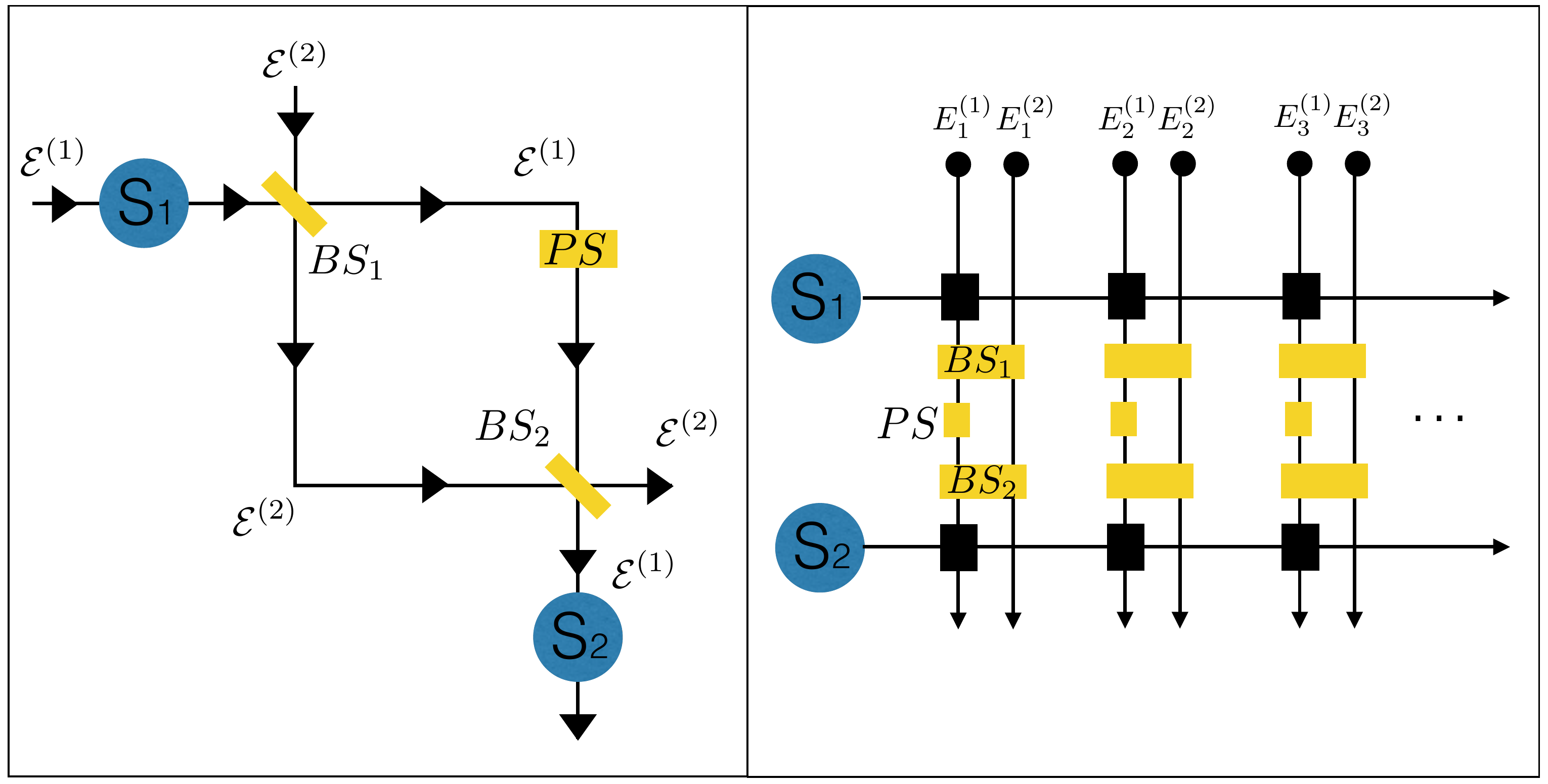}
\caption{Left panel: A sketch of the QCS scheme discussed in Sec.~\ref{sec:interferometer}.  $S_1$ and $S_2$ are two quantum system connected via two unidirectional 
bosonic channels ${\cal E}^{(1)}$ and ${\cal E}^{(2)}$ which are interweaved to form a Mach-Zehnder interferometer ($BS_1$, $BS_2$ being beam splitters and $PS$ being a phase-shifter element).  Right panel: causal flowchart of the couplings of the model in the collisional 
approach. }
\label{fig:interferometer}
\end{figure}

As a first example we analyze  the scheme of Fig.~\ref{fig:interferometer} where $M=2$ quantum systems 
%$M=2$  
%monochromatic quantum electro-dynamical (QED) cavities 
$S_1$ and $S_2$ which can be identified either with monochromatic quantum electro-dynamical (QED) cavities
of  frequency $\omega$ or with two two-level atoms of energy gap $\hbar \omega$, 
interacting 
via $K=2$ unidirectional (chiral) optical channels ${\cal E}^{(1)}$ and ${\cal E}^{(2)}$ that are interweaved to form a Mach-Zehnder interferometer. 
Specifically the environment $\mathcal{E}^{(1)}$, which we assume to be in a thermal state of temperature $T_1$,  interacts  with the first subsystem $S_1$ via a standard 
excitation-hopping term. The output from $S_1$ is then mixed with the second environment $\mathcal{E}^{(2)}$ (initialized at temperature $T_2$)  in a first beam splitter $BS_1$, and then the two signals follow two paths accumulating a phase shift $PS$,  before mixing once again in the second beam splitter $BS_2$. Finally the output from one of the two ports is sent to the second subsystem $S_2$.
%%%%%%%%%%%%%%%%%%%%%
%%%%%%%%%%%%%%%%%%%

In the collisional approach  we shall represent  ${\cal E}^{(1)}$  (${\cal E}^{(2)}$) as  a collection of independent monochromatic  optical quantum carriers $\{ E_{n}^{(1)}; n=1,2,\cdots \}$
(resp. $\{ E_{n}^{(2)}; n=1,2,\cdots \}$)
described   by the annihilation operators $\{ \hat{b}_{E_{n}^{(1)}}; n=1,2 \cdots \}$ (resp. $\{\hat{b}_{E_{n}^{(2)}}; n=1,2 \cdots \}$) each initialized into 
Gibbs states of temperature $T_1$ (resp. $T_2$), i.e. the Gaussian state 
\begin{eqnarray} \label{GAUS1} 
\hat{\eta}_{E^{(1)}_n}:= \frac{\exp[{ - \beta_1  \hat{b}_{E_{n}^{(1)}}^\dag \hat{b}_{E_{n}^{(1)}}}] }{\mbox{Tr}[\exp[{ - \beta_1  \hat{b}_{E_{n}^{(1)}}^\dag \hat{b}_{E_{n}^{(1)}}}] ]},\end{eqnarray}
with 
$\beta_1 = \hbar \omega /k_B T_1$ (resp. $\hat{\eta}_{E^{(2)}_n}$ with 
$\beta_2 = \hbar \omega /k_B T_2$). 
Accordingly the input states $\hat{\eta}_{{\cal E}_n}$ of  Eq.~(\ref{eq:recursive_density_matrix_equation})  are now expressed as 
\begin{eqnarray}  \label{GAUS12} 
\hat{\eta}_{{\cal E}_n} = \hat{\eta}_{E^{(1)}_n} \otimes \hat{\eta}_{E^{(2)}_n}\;. 
\end{eqnarray} 
The interactions between such elements and $S_1$, $S_2$   will follow the causal structure depicted on the right panel of Fig.~\ref{fig:interferometer}. 
In particular we assume no direct couplings between  $\{ \hat{b}_{E_{n}^{(2)}}; n=1,2 \cdots \}$ and the cavities, i.e. 
\begin{eqnarray} \hat{H}_{S_m,{E}^{(2)}_n}=0\;, \label{HAM2} 
\end{eqnarray} 
and take the Hamiltonian~(\ref{COUPH}) which describes the interaction with the  modes  $\{ \hat{b}_{E_{n}^{(1)}}; n=1,2 \cdots \}$ as
\begin{eqnarray}  \label{HAM1} 
\hat{H}_{S_m,{E}^{(1)}_n}= \hat{a}_m^{\dagger}\; \hat{b}_{E_{n}^{(1)}}+\hat{b}_{E_{n}^{(1)}}^\dag\; \hat{a}_m\;,
\end{eqnarray} 
where, for $m=1,2$,  $\hat{a}_m$, $\hat{a}_m^\dag$ are the annihilation  and creation operators of the cavity ${S}_m$, or in case where $S_1$, $S_2$ correspond to 
two-level quantum systems, to the associated lowering and raising Pauli operators. 
Finally we have to specify the structure of the CPT map ${\cal M}_{{\cal E}_n}^{(1)}$ which is responsible for the evolution of the carriers $E_{n}^{(k)}$
between  their collisions with $S_1$ and  $S_2$ (see Fig.~\ref{figure2}) and possibly for the emergence of interference effects in the model. 
In the case we are studying it is given by  the concatenation of three unitary terms,  
 $\hat{V}_{BS_2} \hat{V}_{PS} \hat{V}_{BS_1}$, the first and the third being associated respectively  with the beam-splitter transformations $BS_1$ and $BS_2$  that couple the two channels,
 the second with the phase shift transformation $PS$ acting on the carriers of  ${\cal E}_2$ only, i.e.
 \begin{eqnarray} 
 {\cal M}_{{\cal E}_n}^{(1)}(\cdots) = \hat{V}_{BS_2} \hat{V}_{PS} \hat{V}_{BS_1}( \cdots )V^\dag_{BS_1} \hat{V}_{PS}^\dag \hat{V}_{BS_2}^\dag\;. 
 \end{eqnarray} 
  Specifically, indicating with $\epsilon_j$ the transmissivity of $BS_j$,  the action of  $\hat{V}_{BS_j}$ is fully determined by the identities 
\begin{eqnarray} 
\label{eq:mapbs1}
\hat{V}_{BS_j}^\dag \hat{b}_{E_{n}^{(1)}} \hat{V}_{BS_j} &=& 
\sqrt{\epsilon_j}\, \hat{b}_{E_{n}^{(1)}}-i\sqrt{1-\epsilon_j}\,\hat{b}_{E_{n}^{(2)}}\;,
\\
\label{eq:mapbs2}
\hat{V}_{BS_j}^\dag \hat{b}_{E_{n}^{(2)}} \hat{V}_{BS_j} &=&
-i\sqrt{1-\epsilon_j }\, \hat{b}_{E_{n}^{(1)}}+\sqrt{\epsilon_j}\,\hat{b}_{E_{n}^{(2)}}\;,
\end{eqnarray}
while the action of $\hat{V}_{PS}$ by the identity 
\begin{eqnarray} 
\label{eq:mapps1}
\hat{V}_{PS}^\dag \hat{b}_{E_{n}^{(1)}} \hat{V}_{PS} &=& 
e^{-i\varphi} \hat{b}_{E_{n}^{(1)}}\;,
\\ 
\label{eq:mapps2}
\hat{V}_{PS}^\dag \hat{b}_{E_{n}^{(2)}} \hat{V}_{PS} &=&  \hat{b}_{E_{n}^{(2)}}\;.
\end{eqnarray}
It is worth observing that, in the limit where  $\epsilon_1=\epsilon_2=1$ (i.e. no mixing between ${\cal E}_1$ and ${\cal E}_2$) and $T_1=0$, the model just described reproduce  the one analyzed in Ref.~\cite{1367-2630-14-6-063014}
for $M=2$ two-level atoms.

We first observe that with the above choices the stability condition 
(\ref{eq:stability_condition}) is fulfilled. Indeed from Eq.~(\ref{HAM2}) follows trivially  $\gamma_{m (2)}^{(\ell)}= 0$. Instead 
 from Eq.~(\ref{HAM1}) we can take
   \begin{eqnarray} 
\hat{A}_{S_m}^{(\ell,k)} &=&\delta_{k,1} \left\{ \begin{array}{ccl} 
 \hat{a}_{m}^\dag &&\mbox{for $\ell=1$}  \\ 
\hat{a}_{m} && \mbox{for $\ell=2$,}  \end{array} \right.  \label{QUESTA}  \\
\hat{B}_{S_m}^{(\ell,k)} &=&\delta_{k,1} \left\{ \begin{array}{ccl} 
 \hat{b}_{E_{n}^{(1)}} &&\mbox{for $\ell=1$}  \\ 
\hat{b}^\dag_{E_{n}^{(1)}}&& \mbox{for $\ell=2$,}  \end{array} \right. \label{QUELLA} 
\end{eqnarray} 
so that
 \begin{eqnarray} 
 \gamma_{1 (1)}^{(1)} = [\gamma_{1 (1)}^{(2)}]^* = 
\Big\langle     \hat{b}_{E_{n}^{(1)}}  \; \hat{\eta}_{{\cal E}_n}
\Big\rangle_{{\cal E}} 
= \Big\langle     \hat{b}_{E_{n}^{(1)}}  \hat{\eta}_{E^{(1)}_n} \Big\rangle_{{\cal E}} =0 \;, \nonumber \\\label{eq:stability_condition112} 
\end{eqnarray}
which trivially follow from the fact that the annihilation operator admits zero expectation value on Gibbs states. 
Analogously we have 
\begin{eqnarray} 
\label{eq:stability_condition11}  \gamma_{2 (1)}^{(1)} &=& [\gamma_{2 (1)}^{(2)}]^* = 
\Big\langle     \hat{b}_{E_{n}^{(1)}}  {\cal M}^{(1)}_{{\cal E}_n} ( \hat{\eta}_{{\cal E}_n}) \Big\rangle_{{\cal E}} =\Big\langle   \tilde{\cal M}^{(1)}_{{\cal E}_n} (  \hat{b}_{E_{n}^{(1)}} )\;   \hat{\eta}_{{\cal E}_n} \Big\rangle_{{\cal E}}  \nonumber \\
&=&c(\varphi)   \Big\langle     \hat{b}_{E_{n}^{(1)}} \hat{\eta}_{E^{(1)}_n}  \Big\rangle_{{\cal E}} +  s(\varphi) \Big\langle     \hat{b}_{E_{n}^{(2)}} \hat{\eta}_{E^{(2)}_n}  \Big\rangle_{{\cal E}} =0\;, 
\end{eqnarray}
where  $\tilde{\cal M}^{(1)}_{{\cal E}_n}$ is the complementary counterpart of  ${\cal M}^{(1)}_{{\cal E}_n}$ fulling the property
\begin{eqnarray} 
\tilde{{\cal M}}^{(1)}_{{\cal E}_n} ( \hat{b}_{E_{n}^{(1)}} ) &:=& (V^\dag_{BS_1} \hat{V}_{PS}^\dag \hat{V}_{BS_2}^\dag )   \hat{b}_{E_{n}^{(1)}}     (\hat{V}_{BS_2} \hat{V}_{PS} \hat{V}_{BS_1}) \nonumber \\
&=& c(\varphi) \hat{b}_{E_{n}^{(1)}} + s(\varphi) \hat{b}_{E_{n}^{(2)}}\;,  \end{eqnarray} 
with 
\begin{eqnarray}  
c(\varphi)  &=& e^{-i\varphi}\sqrt{\epsilon_1\epsilon_2} - \sqrt{(1-\epsilon_1) (1-\epsilon_2)}\;,    \nonumber \\
s(\varphi)  &=& -i e^{-i\varphi}\sqrt{(1-\epsilon_1)\epsilon_2} -i \;\sqrt{\epsilon_1 (1-\epsilon_2)}\;. \label{DEFCVAR} 
\end{eqnarray} 
In a similar way we can evaluate the coefficients   $\gamma_{m(kk')}^{(\ell,\ell')}$, $\zeta_{mm'(kk')}^{(\ell,\ell')}$, and $\xi_{mm'(kk')}^{(\ell,\ell')}$
that define the super-operators~(\ref{DEFLLOC}) and (\ref{DEFDLOC}). First of all we notice that 
from  Eq.~(\ref{HAM2}) it follows that only the terms with $k=k'=1$ can have non vanishing values. Next, indicating with 
\begin{eqnarray} 
N_k = (e^{\beta_k}-1)^{-1}\;,
\end{eqnarray} 
the mean photon numbers of the $k$-th thermal bath, 
 we observe that for the local terms of $S_1$  the following identities hold: 
\begin{eqnarray} \nonumber
\gamma_{1(kk')}^{(1,1)} &=& \big[ \gamma_{1(kk')}^{(2,2)} \big]^*= \delta_{k,1}   \delta_{k',1} \;
 \Big\langle  \hat{b}_{E_{n}^{(1)}}^2  
\; \hat{\eta}_{{E}_{n}^{(1)}}   \Big\rangle_{\cal E}  =0 \;, \\
\gamma_{1(kk')}^{(2,1)} &=&  \delta_{k,1}   \delta_{k',1} \;
 \Big\langle  \hat{b}^\dag_{E_{n}^{(1)}}  \hat{b}_{E_{n}^{(1)}}
\; \hat{\eta}_{{E}_{n}^{(1)}}   \Big\rangle_{\cal E}  = \delta_{k,1}   \delta_{k',1} \; N_{1} \;, \nonumber \\
\gamma_{1(kk')}^{(1,2)} &=&  \delta_{k,1}   \delta_{k',1} \;
 \Big\langle  \hat{b}_{E_{n}^{(1)}}  \hat{b}^\dag_{E_{n}^{(1)}} 
\; \hat{\eta}_{{E}_{n}^{(1)}}   \Big\rangle_{\cal E}  = \delta_{k,1}   \delta_{k',1} \; (N_{1} +1)\;,  \nonumber 
\end{eqnarray}
where $\delta_{k,k'}$ indicates the Kronecker delta and where we used known properties of the  second order expectation values of the Gibbs states.  
  Accordingly
    the associated super-operator~(\ref{DEFLLOC}) becomes 
\begin{eqnarray}\label{LOC1ex} 
{\cal L}_1(\cdots)  &=& {({N}_1+1)}\Big(\hat{a}_1(\cdots) \hat{a}_1^{\dagger}-\frac{1}2 \Big[ \hat{a}_1^{\dagger}\hat{a}_1,\cdots \Big]_+\Big) \nonumber \\ &+&
{N}_1\Big(\hat{a}_1^{\dagger}(\cdots) \hat{a}_1-\frac{1}{2}\Big[ \hat{a}_1\hat{a}_1^{\dagger},\cdots \Big]_+\Big)\;,
\end{eqnarray} 
which is already in the standard  GKSL form~(\ref{normalL}) and which  
describes a thermalization process where $S_1$ absorbs and emits excitations from  a thermal bath at temperature $T_1$. 
Similarly the local terms for the $S_2$ gives
\begin{eqnarray} 
&&\gamma_{2 (kk')}^{(1,1)} = \big[ \gamma_{2(kk')}^{(2,2)} \big]^*= \delta_{k,1}   \delta_{k',1} \;
 \Big\langle  \hat{b}^2_{E_{n}^{(1)}} 
\;  {\cal M}^{(1)}_{{\cal E}_n} ( \hat{\eta}_{{\cal E}_n})   \Big\rangle_{\cal E}  \nonumber \\
&&=  \delta_{k,1}   \delta_{k',1} \Big\langle  \left( c(\varphi) \hat{b}_{E_{n}^{(1)}} + s(\varphi) \hat{b}_{E_{n}^{(2)}} \right)^2  \; \hat{\eta}_{{\cal E}_n}   \Big\rangle_{\cal E} =0\;, \end{eqnarray}
and 
\begin{eqnarray} 
\label{eq:coefficient2a}
\gamma_{2(kk')}^{(2,1)} &= & \delta_{k,1}   \delta_{k',1} \;
 \Big\langle \left( c^*(\varphi) \hat{b}^\dag_{E_{n}^{(1)}} + s^*(\varphi) \hat{b}^\dag_{E_{n}^{(2)}} \right)\nonumber \\
 && \times \left( c(\varphi) \hat{b}_{E_{n}^{(1)}} + s(\varphi) \hat{b}_{E_{n}^{(2)}} \right) 
\; \hat{\eta}_{{\cal E}_{n}}   \Big\rangle_{\cal E}  \nonumber \\
&=& \delta_{k,1}   \delta_{k',1} \;N_{12}(\varphi)   \;,
\\
\label{eq:coefficient2b}
 \gamma_{2(kk')}^{(1,2)} &= & \delta_{k,1}   \delta_{k',1} \;
 \Big\langle \left( c(\varphi) \hat{b}_{E_{n}^{(1)}} + s(\varphi) \hat{b}_{E_{n}^{(2)}} \right)\nonumber \\
 && \times \left( c^*(\varphi) \hat{b}^\dag_{E_{n}^{(1)}} + s^*(\varphi) \hat{b}^\dag_{E_{n}^{(2)}} \right) 
\; \hat{\eta}_{{\cal E}_{n}}   \Big\rangle_{\cal E}  \nonumber \\
&=& \delta_{k,1}   \delta_{k',1} \;( N_{12}(\varphi)  + 1)   \;.
\end{eqnarray}
where we introduced
\begin{eqnarray} 
N_{12}(\varphi) &=&  |c(\varphi)|^2  N_1 
+  |s(\varphi)|^2  N_2 \nonumber \\ &=& N_2 + (N_1-N_2)  |c(\varphi)|^2\;,\label{NewNLL} 
\end{eqnarray} 
Replacing all this into Eq.~(\ref{DEFLLOC}) we hence get the following  super-operator
\begin{eqnarray}\label{LOC2ex} 
{\cal L}_2(\cdots)  &=& ({N}_{12}(\varphi)+1)\Big(\hat{a}_2(\cdots) \hat{a}_2^{\dagger}-\frac{1}{2} \Big[ \hat{a}_2^{\dagger}\hat{a}_2,\cdots \Big]_+\Big) \nonumber \\ &+&
{N}_{12} (\varphi)\Big(\hat{a}_2^{\dagger}(\cdots) \hat{a}_2-\frac{1}{2}\Big[ \hat{a}_2\hat{a}_2^{\dagger},\cdots \Big]_+\Big)\;, \end{eqnarray} 
which represents a thermalization process induced by an effective bath whose temperature is intermediate between the one of ${\cal E}_1$ and ${\cal E}_2$ and depends by the 
mixing of the signals induced by their propagation through the Mach-Zehnder.

Consider next the non-local contribution ${\cal D}_{1,2}$ of the master equation. 
In this case we get
\begin{eqnarray} 
&&\zeta_{1,2 (kk')}^{(1,1)}=\left[\xi_{1,2 (kk')}^{(2,2)}\right]^* = \delta_{k,1}   \delta_{k',1}   \Big\langle  \hat{b}_{E_{n}^{(1)}} 
\;  {\cal M}^{(1)}_{{\cal E}_n} (\hat{b}_{E_{n}^{(1)}}  \hat{\eta}_{{\cal E}_n})   \Big\rangle_{\cal E}  \nonumber \\
&& \quad =   \delta_{k,1}   \delta_{k',1}   \Big\langle  ( c(\varphi)  \hat{b}_{E_{n}^{(1)}}  + s(\varphi) \hat{b}_{E_{n}^{(2)}}) 
\; \hat{b}_{E_{n}^{(1)}}  \hat{\eta}_{{\cal E}_n}   \Big\rangle_{\cal E}  =0 \;,  \nonumber  \\
&&\zeta_{1,2 (kk')}^{(2,2)}=\left[\xi_{1,2 (kk')}^{(1,1)}\right]^* =  \delta_{k,1}   \delta_{k',1}   \Big\langle  \hat{b}^\dag_{E_{n}^{(1)}} 
\;  {\cal M}^{(1)}_{{\cal E}_n} (\hat{b}^\dag_{E_{n}^{(1)}}  \hat{\eta}_{{\cal E}_n})   \Big\rangle_{\cal E}  \nonumber \\
&& \quad =     \delta_{k,1}   \delta_{k',1}   \Big\langle  ( c^*(\varphi)  \hat{b}^\dag_{E_{n}^{(1)}}  + s^*(\varphi) \hat{b}^\dag_{E_{n}^{(2)}}) 
\; \hat{b}^\dag_{E_{n}^{(1)}}  \hat{\eta}_{{\cal E}_n}   \Big\rangle_{\cal E}  =0 \;, \nonumber \\ 
\end{eqnarray} 
and 
\begin{eqnarray} 
&&\zeta_{1,2 (kk')}^{(1,2)}=\left[\xi_{1,2 (kk')}^{(2,1)}\right]^* = \delta_{k,1}   \delta_{k',1}   \Big\langle  \hat{b}^\dag_{E_{n}^{(1)}} 
\;  {\cal M}^{(1)}_{{\cal E}_n} (\hat{b}_{E_{n}^{(1)}}  \hat{\eta}_{{\cal E}_n})   \Big\rangle_{\cal E}  \nonumber \\
&& \quad =    \delta_{k,1}   \delta_{k',1}   \Big\langle  ( c^*(\varphi)  \hat{b}^\dag_{E_{n}^{(1)}}  + s^*(\varphi) \hat{b}^\dag_{E_{n}^{(2)}}) 
\; \hat{b}_{E_{n}^{(1)}}  \hat{\eta}_{{\cal E}_n}   \Big\rangle_{\cal E}  \nonumber  \\
&& \quad =    \delta_{k,1}   \delta_{k',1}  \;  c^*(\varphi)\; N_1 \;, \nonumber \\
&& \zeta_{1,2 (kk')}^{(2,1)}=\left[\xi_{2,1 (kk')}^{(1,2)}\right]^* = \delta_{k,1}   \delta_{k',1}   \Big\langle  \hat{b}_{E_{n}^{(1)}} 
\;  {\cal M}^{(1)}_{{\cal E}_n} (\hat{b}^\dag_{E_{n}^{(1)}}  \hat{\eta}_{{\cal E}_n})   \Big\rangle_{\cal E}  \nonumber \\
&& \quad =     \delta_{k,1}   \delta_{k',1}   \Big\langle  ( c(\varphi)  \hat{b}_{E_{n}^{(1)}}  + s(\varphi) \hat{b}_{E_{n}^{(2)}}) 
\; \hat{b}^\dag_{E_{n}^{(1)}}  \hat{\eta}_{{\cal E}_n}   \Big\rangle_{\cal E}  \nonumber  \\
&& \quad =    \delta_{k,1}   \delta_{k',1} \;   c(\varphi) \; (N_1 +1) \;,
\end{eqnarray} 
so that 
\begin{eqnarray}&&
{\cal D}_{1\rightarrow 2}(\cdots)= 
N_1 \Big\{
 c^*(\varphi)
\; \hat{a}_{1}^\dag \Big[\cdots,\hat{a}_2\Big]_{-} - 
 c(\varphi)   
\Big[\cdots, \hat{a}_2^\dag \Big]_{-}\hat{a}_1\Big\} \nonumber 
\\ &&  \quad + \; 
(N_1 +1)  \Big\{  c(\varphi) 
\; \hat{a}_1 \Big[\cdots,\hat{a}_2^\dag \Big]_{-}
-  c^*(\varphi) \;  
\Big[\cdots, \hat{a}_2 \Big]_{-} \hat{a}_{1}^\dag  \Big\} \;. \nonumber \\
\label{DEFDLOC111121}
\end{eqnarray}
One notices that at variance with the contribution~(\ref{LOC1ex}) which fully define the dynamics of $S_1$,  both the local term (\ref{LOC2ex}) of $S_2$ and the coupling super-operator (\ref{DEFDLOC111121}) are modulated 
by the phase $\varphi$. In particular by setting the transmissivities  of  $BS_1$ and $BS_2$  at  $50 \%$ (i.e. $\epsilon_1 =\epsilon_2=0.5$), 
the coefficient $c(\varphi)$ will acquire an oscillating behavior nullifying for $\varphi=\pm \pi$ (specifically  we get
$c(\varphi) = -i e^{-i \varphi/2} \sin(\varphi/2)$). By controlling the parameter $\varphi$ we can hence modify the cascade coupling between $S_1$ and $S_2$. 

Following the derivation of Sec.~\ref{DD} we can finally write  the QCS super-operator in the GKSL form~(\ref{DEFDLOCnew}). In particular in this case the effective Hamiltonian appearing in Eq.~(\ref{DEFDLOC})  is given by
\begin{eqnarray}
\hat{H}_{1,2} &=&- \frac{i}{2} \left( c(\varphi) \;  \hat{a}_2^\dag \hat{a}_1- c^*(\varphi) \; \hat{a}_1^\dag \hat{a}_2 \right)  \label{CHIRHAMMZqwqw}  \\
&=&  - \frac{i}{2} |c(\varphi)| \left(  e^{i \arg[c(\varphi)]}  \hat{a}_2^\dag \hat{a}_1-  e^{-i \arg[c(\varphi)]} \hat{a}_1^\dag \hat{a}_2 \nonumber
\right) \;,
\end{eqnarray}
which by absorbing the phase  $\arg[c(\varphi)]$ into (say) $\hat{a}_1$ exhibits the same chiral symmetry under exchange of $S_1$ and $S_2$ (i.e. $\hat{H}_{2,1} = - \hat{H}_{1,2}$) observed in Ref.~\cite{1367-2630-14-6-063014}. 
The super-operator $\Delta{\cal L}_{1,2}$ of Eq.~(\ref{DEFDLOC}) instead in this case is given by 
\begin{eqnarray} 
&& \Delta{\cal L}_{1,2}(\cdots) = N_1 c^*(\varphi) \left( \hat{a}_1^\dag (\cdots) \hat{a}_2 - \frac{1}{2} \left[ \hat{a}_1^\dag \hat{a}_2, (\cdots) \right]_+\right) \nonumber \\
&& \; + ( N_1 +1) c(\varphi) \left( \hat{a}_1 (\cdots) \hat{a}_2^\dag -\frac{1}{2}  \left[ \hat{a}_2^\dag \hat{a}_1, (\cdots) \right]_+\right) + h.c. \nonumber \\
 \label{CHIRLAZM}
\end{eqnarray}
which, remembering~(\ref{QUESTA}), can be expressed as in (\ref{CHIRLA}) with
\begin{eqnarray} \Delta D_{m_1,m_2(k_1k_2)}^{(\ell_1,\ell_2)}=\delta_{k,1}   \delta_{k',1}   \Delta D_{m_1,m_2(1,1)}^{(\ell_1,\ell_2)}\;, 
\end{eqnarray} 
where for $m_1,m_2=1,2$ and $\ell_1,\ell_2=1,2$, $\Delta D_{m_1,m_2(1,1)}^{(\ell_1,\ell_2)}$ is the $4\times 4$ matrix of elements
\begin{eqnarray} 
\left[\begin{array}{cc|cc}
0 & 0& N_1  c^*(\varphi) &0  \\
0 & 0 & 0 &(N_1 +1)c(\varphi)  \\\hline 
N_1c(\varphi) & 0 & 0 & 0 \\ 
0 & (N_1+1)c^*(\varphi) &0&0  \end{array}\right] \nonumber 
\end{eqnarray}
the top-left and bottom right $2\times 2$ blocks being associated with $m_1=m_2= 1$ and $m_1 = m_2=2$ respectively. 
As anticipated in the previous section, while being Hermitian,  this is in general not positive semi-definite (indeed it admits eigenvalues 
$\pm N_1 |c(\varphi)|$ and $\pm (N_1 +1) |c(\varphi)|$).
On the contrary the matrix (\ref{COEFMEW}) which describe the sum of  $\Delta{\cal L}_{1,2}$ with the local terms ${\cal L}_{1}$ of Eq.~(\ref{LOC1ex}) and ${\cal L}_2$  of Eq.~(\ref{NewNLL})  is given by 
\begin{eqnarray} 
\left[\begin{array}{cc|cc}
N_1 & 0& N_1  c^*(\varphi) &0  \\
0 & N_1+1 & 0 &(N_1 +1)c(\varphi)  \\\hline 
N_1c(\varphi) & 0 & N_{12}(\varphi)  & 0 \\ 
0 & (N_1+1)c^*(\varphi) &0&N_{12}(\varphi) +1  \end{array}\right] \nonumber \\
\nonumber
\end{eqnarray}
and has eigenvalues
\begin{eqnarray}
\kappa_{1,\pm} &=&   \frac{1}{2} \Big(  N_1 + N_{12}(\varphi) +2  \\
&&  \pm \sqrt{(N_1 - N_{12})^2 + 4(N_1+1)^2  |c(\varphi)|^2}\Big) \;, \nonumber  \\
\kappa_{2,\pm} &=&  \frac{1}{2} \Big( N_1 + N_{12}(\varphi)\\
&&  \pm \sqrt{(N_1 - N_{12})^2 + 4N_1^2  |c(\varphi)|^2} \Big)\;, \nonumber 
\end{eqnarray} 
which are  non-negative for all possible choices of $N_1, N_2\geq 0$ and $|c(\varphi)|\in [0,1]$. 
The associated Lindblad operators~\eqref{LLLLLLL} can instead be shown to be equal to 
\begin{eqnarray}
\hat{L}^{(1,+)}&=&\sqrt{k_{1,+}} \; \frac{w_{1,+}\; \hat{a}_1 + \hat{a}_2}{\sqrt{1+|w_{1,+}|^2}} \;, \\ 
\hat{L}^{(1,-)}&=&\sqrt{k_{1,-}}\; \frac{w_{1,-} \; \hat{a}_1^{\dagger}+ \hat{a}_2^{\dagger}}{\sqrt{1+|w_{1,-}|^2}}\;, \\
\hat{L}^{(2,+)}&=&\sqrt{k_{2,+}}\; \frac{w_{2,+}\; \hat{a}_1+  \hat{a}_2}{\sqrt{1+|w_{2,+}|^2}} \;, \\
\hat{L}^{(2,-)}&=&\sqrt{k_{2,-}}\; \frac{w_{2,-}  \; \hat{a}_1^{\dagger} +   \hat{a}_2^{\dagger} }{\sqrt{1+|w_{2,-}|^2}}\;,
\end{eqnarray}
with
\begin{eqnarray}
w_{1,\pm}&=&\frac{1}{{2(N_1+1)c^*(\varphi)}} \Big[ N_1-N_{12}(\varphi)  \\\nonumber
&&\pm\sqrt{(N_1-N_{12}(\varphi))^2+4(N_1+1)^2|c(\varphi)|^2} \Big]\;, \\ 
%{2(N_1+1)c^*(\varphi)}\;,\\
w_{2,\pm}&=&\frac{1}{2N_1c(\varphi)} \Big[ 
N_1-N_{12}(\varphi) \\ \nonumber
&&\pm\sqrt{(N_1-N_{12}(\varphi))^2+4N_1^2|c(\varphi)|^2} \Big]\;.
\end{eqnarray}
It is worth noticing that in the already cited limit of $\epsilon_{1,2}=1$ and $T_1=0$ reproducing the model in~\cite{1367-2630-14-6-063014}, we have that only the eigenvalue $k_{1,+}=2$ is different from zero, so that one
has only one collective jump operator
$\hat{L}^{(1,+)}=\hat{a}_1+\hat{a}_2$.

\subsection{Example 2: controlling the topology of the network via interference }
\label{sec:example2}
%%%%%%%
\begin{figure}[!t]
\centering
\includegraphics[scale=0.25]{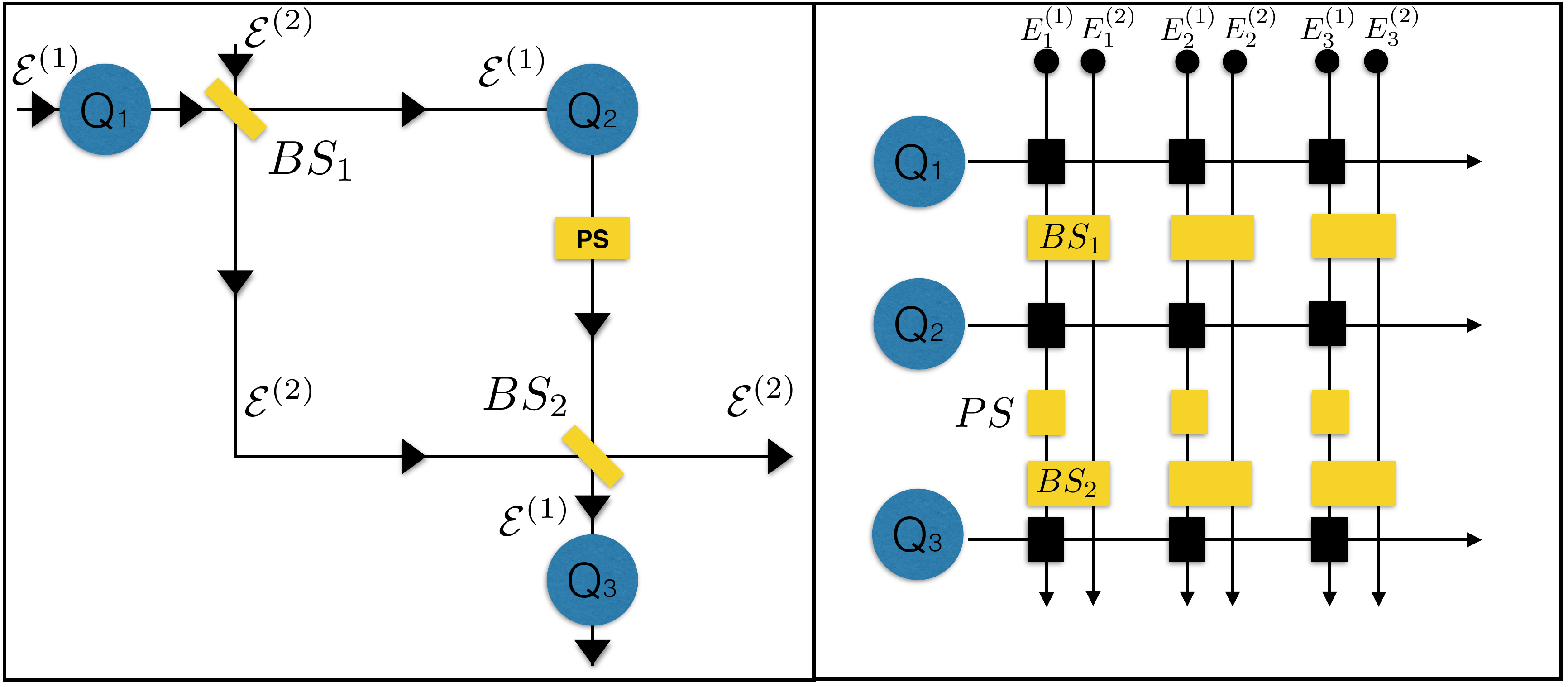}
\caption{Left panel: A sketch of the QCS scheme discussed in Sec~\ref{sec:example2}. $Q_1$, $Q_2$, $Q_3$ are the quantum system elements which are connected 
by the QCS network formed by the unidirectional bosonic channels 
$\mathcal{E}^{(1)}$ and $\mathcal{E}^{(2)}$. As in the case of Sec.~\ref{sec:interferometer} they
are  interweaved by two beam splitters and a phase shifter. Right panel: causal flowchart of the couplings of the system in the collisional model.}
\label{fig:example2}
\end{figure}
%%%%%%

In this section we discuss how interference can be used to effectively modify the topology of the QCS interaction network  by selectively activating/deactivating some of the couplings
which enter the scheme.  In particular we focus on the case of 
 three quantum systems, dubbed  $Q_1$, $Q_2$ and $Q_3$ 
 connected as  schematically shown   in Fig.~\ref{fig:example2}. This is basically the same
 configuration discussed in Sec.~\ref{sec:interferometer} where $Q_1$ and $Q_3$ take the positions of $S_1$ and $S_2$ respectively, while 
 $Q_2$ is placed inside the Mach-Zehnder interferometer.   Accordingly the model exhibits direct QCS connections among first neighboring elements (i.e. the couple  $Q_1$ and  $Q_2$ and 
 the couple $Q_2$ and $Q_3$), while the QCS coupling among $Q_1$ and $Q_3$ is mediated by two channels which interfere. 
The dynamics of the model can be derived 
following the same line of the previous section -- see Appendix~\ref{appendixfin} for the explicit calculations. 

Expressed as in Eq.~(\ref{DEFC}) the resulting master equation exhibits the following local contributions:
\begin{eqnarray}\label{LOC1exnew} 
{\cal L}_1(\cdots)  &=& {({N}_1+1)}\Big(\hat{a}_1(\cdots) \hat{a}_1^{\dagger}-\frac{1}2 \Big[ \hat{a}_1^{\dagger}\hat{a}_1,\cdots \Big]_+\Big) \nonumber \\ &+&
{N}_1\Big(\hat{a}_1^{\dagger}(\cdots) \hat{a}_1-\frac{1}{2}\Big[ \hat{a}_1\hat{a}_1^{\dagger},\cdots \Big]_+\Big)\;, \\
\mathcal{L}_2(\cdots)&=& \Big(\bar{N}_{12}+1\Big) \; \Big(\hat{a}_2(\cdots)\hat{a}_2^{\dagger}-\frac{1}{2}\Big[\hat{a}_2^{\dagger}\hat{a}_2,\cdots\Big]_+\Big) \nonumber \\
&+&\bar{N}_{12}\; \Big(\hat{a}_2^{\dagger}(\cdots)\hat{a}_2-\frac{1}{2}\Big[\hat{a}_2\hat{a}_2^{\dagger},\cdots\Big]_+\Big)\;,  \label{DEFL2new}\\
\mathcal{L}_3(\cdots)&=&\Big(N_{12}(\varphi)+1\Big)\Big(\hat{a}_3(\cdots)\hat{a}_3^{\dagger}-\frac{1}{2}\Big[\hat{a}_3^{\dagger}\hat{a}_3,\cdots\Big]_+\Big) \nonumber \\
&+&N_{12}(\varphi)\Big(\hat{a}_3^{\dagger}(\cdots)\hat{a}_3-\frac{1}{2}\Big[\hat{a}_3\hat{a}_3^{\dagger},\cdots\Big]_+\Big)\;, \label{DEFL3}
\end{eqnarray}
with $N_{12}(\varphi)$ defined as  in  Eq.~(\ref{NewNLL})  and $\bar{N}_{12}$ being the average photon number of the environments perceived by  $Q_2$, i.e. 
 \begin{eqnarray}  \label{DEFNBAR12} 
\bar{N}_{12} = \epsilon_1N_1+(1-\epsilon_1)N_2 = N_2 + \epsilon_1 (N_1 -N_2) \;.
\end{eqnarray} 
Notice that the local terms of $Q_1$  and $Q_3$ coincide respectively with those of $S_1$ and $S_2$ of the previous section and the ${\cal L}_2$ doesn't depend upon the phase
$\varphi$. 

The non-local contributions of the model are instead given by two first-neighboring elements, connecting the couples $Q_1,Q_2$ and $Q_2Q_3$, plus a 
second-neighboring contribution, connecting $Q_1$ and $Q_3$. The first two are given by 
\begin{eqnarray}
\nonumber
&&\mathcal{D}_{1\rightarrow2}(\cdots)=\sqrt{\epsilon_1}N_1\Big(\hat{a}_1^{\dagger}\Big[\cdots,\hat{a}_2\Big]_-+\Big[\hat{a}_2^{\dagger},\cdots \Big]_-\hat{a}_1\Big)\\
&&+ \sqrt{\epsilon_1}\Big(N_1+1\Big)\Big(\hat{a}_1\Big[\cdots,\hat{a}_2^{\dagger}\Big]_-+\Big[\hat{a}_2,\cdots\Big]_-\hat{a}_1^{\dagger}\Big)  \;, \label{DDEFD12}
\end{eqnarray}
and 
\begin{eqnarray}
\mathcal{D}_{2\rightarrow3}(\cdots)&=&M^*_{12}(\varphi) 
\hat{a}_2^{\dagger}\Big[\cdots,\hat{a}_3\Big]_{-} +M_{12}(\varphi)\Big[\hat{a}_3^{\dagger},\cdots\Big]_{-} \hat{a}_2  \nonumber \\
&+& (M_{12}(\varphi) +\lambda(\varphi))
\hat{a}_2\Big[\cdots,\hat{a}_3^{\dagger}\Big]_{-}  \nonumber \\
&+& (M_{12}^*(\varphi) +\lambda^*(\varphi))\Big[\hat{a}_3,\cdots\Big]_{-} \hat{a}_2^{\dagger}\;,\label{DDEFD23} 
\end{eqnarray}
where we introduced the functions 
\begin{eqnarray}
 M_{12}(\varphi)&=&\sqrt{\epsilon_1}c(\varphi)N_1+i\sqrt{1-\epsilon_1}s(\varphi)N_2\;, \nonumber \\
\lambda(\varphi) &=&  \sqrt{\epsilon_1} c(\varphi) +i\sqrt{1-\epsilon_1}s(\varphi)\;,  \label{DEFDEFDEF}
\end{eqnarray}
with $c(\varphi)$ and $s(\varphi)$ as in Eq.~(\ref{DEFCVAR}). The third term instead is given by 
\begin{eqnarray}&&
{\cal D}_{1\rightarrow 3}(\cdots)= 
N_1 \Big\{
 c^*(\varphi)
\; \hat{a}_{1}^\dag \Big[\cdots,\hat{a}_3\Big]_{-} +
 c(\varphi)   
\Big[\hat{a}_3^\dag, \cdots  \Big]_{-}\hat{a}_1\Big\} \nonumber 
\\ &&  \quad + \; 
(N_1 +1)  \Big\{  c(\varphi) 
\; \hat{a}_1 \Big[\cdots,\hat{a}_3^\dag \Big]_{-}
+ c^*(\varphi) \;  
\Big[\hat{a}_3, \cdots \Big]_{-} \hat{a}_{1}^\dag  \Big\} \;, \nonumber \\ \label{DEFD13}
\end{eqnarray}
and formally coincides with the element ${\cal D}_{1\rightarrow 2}(\cdots)$ of the previous section which connected $S_1$ and $S_2$.
The above expressions make it clear that the various coupling terms have different functional dependences upon the phase parameter $\varphi$. 
To better appreciate this it is useful to focus on the  zero temperature regime 
(i.e. $N_1=N_2=0$), and to assume the beam splitters to have $50\%$ transmissivities (i.e. $\epsilon_1=\epsilon_2=1/2$). Under these assumptions
 all the local contributions describe a purely dissipative evolution which is independent from  $\varphi$, i.e. 
 \begin{eqnarray}\label{LOC1exnewEEEE} 
{\cal L}_m (\cdots)  &=&\hat{a}_m(\cdots) \hat{a}_m^{\dagger}-\frac{1}2 \Big[ \hat{a}_m^{\dagger}\hat{a}_m,\cdots \Big]_+ \;, \quad m=1,2,3 \nonumber  \end{eqnarray} 
 while 
   Eqs.~(\ref{DDEFD12}) -- (\ref{DEFD13}) yield
\begin{eqnarray}
\nonumber
\mathcal{D}_{1\rightarrow2}(\dots)&=&\frac{1}{\sqrt{2}}\Big(\hat{a}_1\Big[\cdots,\hat{a}_2^\dagger\Big]_{-}+\Big[\hat{a}_2,\cdots\Big]_{-}\hat{a}_1^\dagger\Big)\;,\\
\nonumber
\mathcal{D}_{2\rightarrow3}(\dots)&=&\frac{1}{\sqrt{2}}\Big( e^{-i \varphi}  \hat{a}_2\Big[\cdots,\hat{a}_3^\dagger\Big]_{-}+ e^{i \varphi}  \Big[\hat{a}_3,\cdots\Big]_{-}\hat{a}_2^\dagger\Big)\;, \\
\nonumber
\mathcal{D}_{1\rightarrow3}(\dots)&=&- i \sin\tfrac{\varphi}{2} \Big( e^{-i \tfrac{\varphi}{2}}  \hat{a}_1\Big[\cdots,\hat{a}_3^\dagger\Big]_{-} \nonumber \\
 &&\qquad \qquad \qquad \qquad +  e^{i \tfrac{\varphi}{2}}  \Big[\hat{a}_3,\cdots\Big]_{-}\hat{a}_1^\dagger\Big). 
\end{eqnarray}
The above equations make it explicit that the parameter $\varphi$ contributes  to the system dynamics in two different ways. First 
 it introduces a non-trivial relative phase between $Q_1$, $Q_2$ and $Q_3$ which, at variance with the two body problem of the previous section cannot be removed by simply redefining their corresponding annihilation/creation operators.  Second it  
induces a  selective modulation of the intensity of the $Q_1Q_3$ interactions. 
%%%%%%%%
\begin{figure}[!t]
\centering
\includegraphics[scale=0.24]{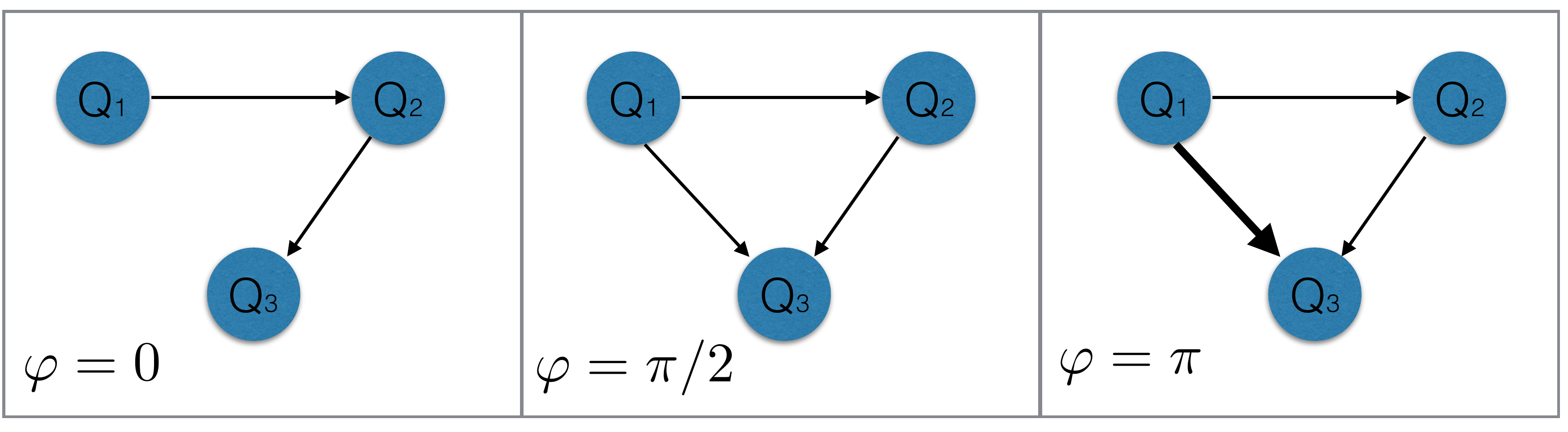}
\caption{Pictorial representation of the QCS  interactions among $Q_1$, $Q_2$ and $Q_3$. Left panel: interaction scheme for $\varphi=0$, where there are only interactions between first neighbors. Central panel: for $\varphi=\pi/2$, $Q_1$ interacts also with $Q_3$ and their interaction is of the same strength as the first neighbor ones. Right panel: for $\varphi=\pi$, not only there is an interaction between $Q_1$ and $Q_3$, but it is even stronger than the first neighbor ones.}
\label{fig:int_top}
\end{figure}
These facts are reflected  into the structure of  the effective Hamiltonian~(\ref{EFFEHAMI}) stemming from the the reshaping of the ME in Lindblad form, i.e. \begin{eqnarray}
\hat{H}_{1,2}&=&-\frac{i}{2\sqrt{2}}\Big(\hat{a}_1\hat{a}_2^{\dagger}-\hat{a}_1^{\dagger}\hat{a}_2\Big)\;, \\
\hat{H}_{2,3}&=&-\frac{i}{2\sqrt{2}}\Big(e^{i\frac{\varphi}{2}} \hat{a}_2\hat{a}_3^{\dagger}-e^{-i\frac{\varphi}{2}} \hat{a}_2^{\dagger}\hat{a}_3\Big)\;, \\
\hat{H}_{1,3}&=&-\frac{i}{2}\sin{\frac{\varphi}{2}}\Big(e^{i\frac{\varphi+\pi}{2}}\hat{a}_1\hat{a}_3^{\dagger}-e^{-i\frac{\varphi+\pi}{2}}\hat{a}_1^{\dagger}\hat{a}_3\Big)\;,
\end{eqnarray}
see Eqs.~(\ref{EFFE1}) -- (\ref{EFFE3}) of Appendix~\ref{appendixfin}. 
Accordingly we see that acting on $\varphi$ the topology of the system interactions can be modified, moving from the case where the  interactions among $Q_1$ and $Q_3$ 
is null  (e.g. $\varphi =0$) or amplified  ($\varphi =\pi$)  with respect to their  $Q_1Q_2$ and $Q_2Q_3$ counterparts, whose associated intensities are instead independent from $\varphi$ -- see Fig.~\ref{fig:int_top}.

\section{Conclusions} \label{SEC:CON}
In this work we developed a general theoretical framework for modeling complex networks of quantum systems organized in a cascade fashion, {\it i.e.}\ such that the coupling between the various subsystems is mediated by unidirectional environmental  channels. Differently from previous approaches, our framework allows also to consider interactions and interference effects between environmental channels, inducing a rich and complex effective dynamics on the nodes of the network.

The theoretical derivation is based on a collisional model that allows to derive a many-body master equation which preserves the positivity of the density matrix and correctly incorporates the causal structure of the network. Moreover, expressing the master equation in Lindblad form, we obtain an effective Hamiltonian coupling between the systems which is externally tunable by properly modifying the parameters of the network. 

We focused on two particular examples: a cascade system in a Mach-Zehnder-like configuration showing dissipative interference effects, and a tripartite cascade network  where the topology of the interactions is controllable by means of a simple phase shifter.
More generally, the possibility of engineering Hamiltonian and dissipative interactions exploiting interference effects in cascade systems is very intriguing and worth to be further investigated in future works.

\appendix
\section{Derivation of the master equation\label{sec:derivation}}
The second order expansion of  Eq.~(\ref{SUPU}) with respect to the product $g\Delta t$ is
\begin{eqnarray}
&& \mathcal{U}_{S_m,{\cal E}_{n}}=\mathcal{I}_{S_m,{\cal E}_{n}}+(g\Delta t)\; \mathcal{U}_{S_m,{\cal E}_{n}}' +(g\Delta t)^2\; \mathcal{U}_{S_m,{\cal E}_{n}}''
\nonumber \\
 && \qquad \qquad \qquad \qquad \qquad \qquad \quad  \qquad+\mathcal{O}(g\Delta t)^3\;, 
\end{eqnarray}
with $\mathcal{I}_{S_m,{\cal E}_{n}}$ being the identity super-operator and 
\begin{eqnarray}
\mathcal{U}_{S_m,{\cal E}_{n}}'(\cdots)&=&-i\sum_{k=1}^K \Big[ \hat{H}_{S_m,{ E}_n^{(k)}},(\cdots)\Big]_-\;, \\
\mathcal{U}_{S_m,{\cal E}_{n}}''(\cdots)&=& \sum_{k,k'=1}^M \; \Big\{  H_{S_m,E^{(k)}_{n}}(\cdots)H_{S_m,E^{(k')}_{n}} \\ \nonumber 
&& \; -\frac{1}{2}\Big[ H_{S_m,E^{(k)}_{n}}\; H_{S_m,E^{(k')}_{n}},(\cdots) \Big]_+ \Big\}.
\end{eqnarray} 
By replacing these expressions into  Eq.~(\ref{DEFC}) we then  obtain the expansion of the super-operator $\mathcal{C}_{\mathcal{S},{\cal E}_{n}}$,i.e. 
  \begin{eqnarray}
\mathcal{C}_{\mathcal{S},{\cal E}_{n}} =\mathcal{C}_{\mathcal{S},{\cal E}_{n}}^0 + (g\Delta t)\; \mathcal{C}_{\mathcal{S},{\cal E}_{n}}' + 
(g\Delta t)^2 \; \mathcal{C}_{\mathcal{S},{\cal E}_{n}}'' \nonumber +\mathcal{O}(g\Delta t)^3\;,
\\ \label{DEFCAPP} 
\end{eqnarray}
where 
\begin{eqnarray}
\mathcal{C}_{\mathcal{S},{\cal E}_{n}}^0&=&  {\cal M}^{(M \leftarrow 1)}_{{\cal E}_n}
,\nonumber \\
\mathcal{C}_{\mathcal{S},{\cal E}_{n}}'&=&\sum_{m=1}^M  {\cal M}^{(M \leftarrow m)}_{{\cal E}_n} 
 \circ \mathcal{U}_{S_m,{\cal E}_{n}}' \circ  
{\cal M}^{(m-1 \leftarrow 1)}_{{\cal E}_n}
, \nonumber 
\\
\mathcal{C}_{\mathcal{S},{\cal E}_{n}}''&=&\mathcal{C}_{\mathcal{S},{\cal E}_{n}}^{''(a)}+\mathcal{C}_{\mathcal{S},{\cal E}_{n}}^{''(b)},
\end{eqnarray} 
and 
\begin{eqnarray} 
\mathcal{C}_{\mathcal{S},{\cal E}_{n}}^{''(a)}&=&
\sum_{m=1}^M
{\cal M}^{(M \leftarrow m)}_{{\cal E}_n}
\circ \mathcal{U}_{S_m,{\cal E}_{n}}'' \circ  
{\cal M}^{(m-1 \leftarrow 1)}_{{\cal E}_n}
, \nonumber  \\
\mathcal{C}_{\mathcal{S},{\cal E}_{n}}^{''(b)}&=&\sum_{m'=m+1}^{M}\sum_{m=1}^{M-1} \Big\{ 
{\cal M}^{(M \leftarrow m')}_{{\cal E}_n}
\nonumber 
\circ\mathcal{U}_{S_{m'},{\cal E}_{n}}' \\ && \circ 
{\cal M}^{(m'-1 \leftarrow m)}_{{\cal E}_n}
\circ \mathcal{U}_{S_m,{\cal E}_{n}}'\circ 
{\cal M}^{(m-1 \leftarrow 1)}_{{\cal E}_n}
\Big\} ,\nonumber \\
\end{eqnarray} 
where we defined 
\begin{eqnarray} \label{DEFMMM}  
{\cal M}^{(m_2 \leftarrow m_1)}_{{\cal E}_n} := \left\{ \begin{array}{ll}
 \overleftarrow{\Pi}_{m=m_1}^{m_2} 
\mathcal{M}^{(m)}_{{\cal E}_{n}}& \mbox{for $m_2 \geq m_1$} \;, \\ \\
{\cal I} & \mbox{for $m_2 <  m_1$}  \;,
\end{array} \right.
\end{eqnarray} 
to indicate the ordered product of the maps $\mathcal{M}^{(m_1)}_{{\cal E}_{n}}$, $\mathcal{M}^{(m_1+1)}_{{\cal E}_{n}}$, $\cdots$, $\mathcal{M}^{(m_2)}_{{\cal E}_{n}}$ -- see also definition~(\ref{SHORT}). 
Inserting all this into  Eq.~(\ref{eq:recursive_density_matrix_equation}) and taking the partial trace with respect to the carriers then allows us to write 
the following equation \begin{eqnarray}\label{RHOAPPR1}
&&\frac{\hat{\rho}(n+1)-\hat{\rho}(n)}{\Delta t} = g \; \Big\langle \mathcal{C}_{\mathcal{S},{\cal E}_{n+1}}'(\hat{R}(n)\otimes \hat{\eta}_{{\cal E}_{n+1}})\Big\rangle_{\cal E}   \nonumber\\
&&\; +g^2\Delta t \;\Big\langle \mathcal{C}_{\mathcal{S},{\cal E}_{n+1}}''(\hat{R}(n)\otimes \hat{\eta}_{{\cal E}_{n+1}})\Big\rangle_{\cal E}+\mathcal{O}(g^3\Delta t^2),
\end{eqnarray}
which by explicit evaluation of the various terms reduces to Eq.~(\ref{RHOAPPR}) of the main text. 
Indeed the first order term in $g$ of this expression can be written as 
\begin{eqnarray}
&& \Big\langle \mathcal{C}_{\mathcal{S},{\cal E}_{n+1}}'(\hat{R}(n)\otimes \hat{\eta}_{{\cal E}_{n+1}})\Big\rangle_{\cal E} 
\\&=&-i\sum_{m,k,\ell}\Big\langle \hat{B}_{E_{n+1}^{(k)}}^{(\ell,m)}
{\cal M}^{(m-1 \leftarrow 1)}_{{\cal E}_{n+1}}
(\hat{\eta}_{{\cal E}_{n+1}})\Big\rangle_{\cal E} \;  \Big[\hat{A}_{S_m}^{(\ell,k)},\hat{\rho}(n)\Big]_{-}, \nonumber 
\end{eqnarray}
and coincides with the first order contribution  of Eq.~(\ref{RHOAPPR}) with 
\begin{eqnarray} \label{LAMBDA1} 
\gamma_{m (k)}^{(\ell)}  = \Big\langle \hat{B}_{E_{n+1}^{(k)}}^{(\ell,m)}
{\cal M}^{(m-1 \leftarrow 1)}_{{\cal E}_{n+1}}
(\hat{\eta}_{{\cal E}_{n+1}})\Big\rangle_{\cal E} \;.
\end{eqnarray} 
Similarly the second order term of (\ref{RHOAPPR1}) is given by two contributions: 
\begin{widetext} 
\begin{eqnarray}
&&
\Big\langle\mathcal{C}_{\mathcal{S},{\cal E}_{n+1}}^{''(a)}(\hat{R}(n)\otimes \hat{\eta}_{{\cal E}_{n+1}})\Big\rangle_{\mathcal{E}}= 
\frac{1}{2} \sum_{m=1}^M \sum_{k,k'}\sum_{\ell,\ell'}   \gamma_{m(kk')}^{(\ell,\ell')} 
 \;\Big\{  2  \hat{A}_{S_m}^{(\ell,k)} \hat{\rho}(n) \hat{A}_{S_m}^{(\ell',k')}
-\Big[ \hat{A}_{S_m}^{(\ell',k')} \hat{A}_{S_m}^{(\ell,k)}  ,\hat{\rho}(n) \Big]_{+} \Big\} \\
&&
\Big\langle\mathcal{C}_{\mathcal{S},{\cal E}_{n+1}}^{''(b)}(\hat{R}(n) \otimes  \hat{\eta}_{{\cal E}_{n+1}} )\Big\rangle_{\mathcal{E}} 
=\sum_{m'=m+1}^{M}\sum_{m=1}^{M-1} 
\sum_{k,k'} \sum_{\ell,\ell'}
\Big\{
\zeta_{mm'(kk')}^{(\ell,\ell')} 
\; \hat{A}_{S_m}^{(\ell,k)}\Big[\hat{\rho}(n),\hat{A}_{S_{m'}}^{(\ell',k')}\Big]_{-}
- \xi_{mm'(kk')}^{(\ell,\ell')} \; 
\Big[\hat{\rho}(n),{\hat{A}_{S_{m'}}^{(\ell',k')}} \Big]_{-}{\hat{A}_{S_m}^{(\ell,k)}} \Big\} \nonumber 
\end{eqnarray}
with 
 coefficients 
 \begin{eqnarray}
\label{eq:general_local_coefficient}
\gamma_{m(kk')}^{(\ell,\ell')} &=&  \Big\langle  \hat{B}_{E_{n+1}^{(k')}}^{(\ell',m)}   \hat{B}_{E_{n+1}^{(k)}}^{(\ell,m)} 
{\cal M}^{(m-1 \leftarrow 1)}_{{\cal E}_{n+1}}
(\hat{\eta}_{{\cal E}_{n+1}})   \Big\rangle_{\cal E}\;, \\
\label{eq:general_interaction_coefficient}
\zeta_{mm'(kk')}^{(\ell,\ell')}&=&
\Big\langle \hat{B}_{E_{n+1}^{(k')}}^{(\ell',m')}
 {\cal M}^{(m'-1 \leftarrow m)}_{{\cal E}_{n+1}}
\Big(\hat{B}_{E_{n+1}^{(k)}}^{(\ell,m)}
 {\cal M}^{(m-1 \leftarrow 1)}_{{\cal E}_{n+1}}
(\hat{\eta}_{{\cal E}_{n+1}})\Big)\Big\rangle_{\mathcal{E}} \;,\\
\xi_{mm'(kk')}^{(\ell,\ell')}&=&  \label{eq:general_interaction_coefficient1}
\Big\langle \hat{B}_{E_{n+1}^{(k')}}^{(\ell',m')}
 {\cal M}^{(m'-1 \leftarrow m)}_{{\cal E}_{n+1}}
\Big(
 {\cal M}^{(m-1 \leftarrow 1)}_{{\cal E}_{n+1}}
(\hat{\eta}_{{\cal E}_{n+1}}) \hat{B}_{E_{n+1}^{(k)}}^{(\ell,m)} \Big)\Big\rangle_{\mathcal{E}} \;.
\end{eqnarray} 
\end{widetext} 

\section{Positivity of the matrix $D_{m_1,m_2(k_1k_2)}^{(\ell_1,\ell_2)}$} \label{APPENDIXb} 
As anticipated in Sec.~\ref{DD} one can show that  the matrix $\Omega$ of elements $\Omega_{j,j'} =D_{m_1,m_2(k_1k_2)}^{(\ell_1,\ell_2)}$  ($j$ being  the joint index $(\ell,k,m)$ and $D_{m_1,m_2(k_1k_2)}^{(\ell_1,\ell_2)}$ as in
Eq.~(\ref{COEFMEW})) is non-negative, i.e. that  for all row vectors $\vec{q}$ of complex elements $q_j$ the following inequality applies 
\begin{eqnarray} 
\vec{q}\;   \Omega\;  \vec{q}^{\;\dag}  : =  \sum_{j,j'}  q_j \; \Omega_{j,j'} \; q_{j'}^*   \geq 0\;.
\end{eqnarray} 
Indeed from Eq.~(\ref{eq:general_interaction_coefficient})-(\ref{eq:general_interaction_coefficient1}) it follows that 
\begin{widetext} 
\begin{eqnarray}
&& 2 \;  \vec{q}\;   \Omega\;  \vec{q}^{\;\dag} =   \sum_{m} q_{(\ell,k,m)} q_{(\ell',k',m)}^*   \gamma_{m(kk')}^{(\ell,\ell')} 
 + \sum_{m'>m} \left[ q_{(\ell,k,m)}  q_{(\ell',k',m')}^*
\zeta_{mm'(kk')}^{(\ell,\ell')}  + {h.c.}  \right]  \label{impoimpo}   \\
&&\quad = \sum_{m} \Big\langle  \hat{Q}_{{\cal E}_{n+1}}^{(m) \dag}   \hat{Q}_{{\cal E}_{n+1}}^{(m)} 
{\cal M}^{(m-1 \leftarrow 1)}_{{\cal E}_{n+1}}
(\hat{\eta}_{{\cal E}_{n+1}})   \Big\rangle_{\cal E}  +  \sum_{m'>m} \left[ \Big\langle
\hat{Q}_{{\cal E}_{n+1}}^{(m') \dag} 
 {\cal M}^{(m'-1 \leftarrow m)}_{{\cal E}_{n+1}}
\Big( \hat{Q}_{{\cal E}_{n+1}}^{(m)}
 {\cal M}^{(m-1 \leftarrow 1)}_{{\cal E}_{n+1}}
(\hat{\eta}_{{\cal E}_{n+1}})\Big)\Big\rangle_{\mathcal{E}} + h.c. \right]
\;, \nonumber 
\end{eqnarray}\end{widetext}  
where in the first line we use 
Eq.~(\ref{fffd1}) and, for the easy of notation,  the convention of sum over repeated indexes, while in the second line we introduce the operators 
\begin{eqnarray} \hat{Q}_{{\cal E}_{n+1}}^{(m)}  =  \sum_{\ell,k} q_{(\ell,k,m)} \hat{B}_{E_{n+1}^{(k)}}^{(\ell,m)}\;.
\end{eqnarray} 
To proceed further we invoke the Stinespring decomposition~\cite{HolevoBOOK}  to write 
\begin{eqnarray} 
\mathcal{M}^{(m)}_{{\cal E}_{n}}(\cdots) &=& \mbox{Tr}_{{\cal A}_n} [ {\cal V}^{(m)}_{{\cal E}_{n}{\cal A}}( \cdots \otimes |0\rangle_{{\cal A}}  \langle 0|) ]\;, 
\\ 
{\cal V}^{(m)}_{{\cal E}_{n}{\cal A}}(\cdots)  &:=&  V^{(m)}_{{\cal E}_{n}{\cal A}}( \cdots ) V^{(m)^\dag}_{{\cal E}_{n}{\cal A}}\;, 
\end{eqnarray} 
with $|0\rangle_{{\cal A}}$ being a (fixed)  reference state of an ancillary system ${\cal A}$ and $V^{(m)}_{{\cal E}_{n}{\cal A}}$ being a  unitary transformation
that couples it with ${\cal E}_{n}$. Accordingly from Eq.~(\ref{DEFMMM}) it follows that for all $m_2 \geq m_1$ one has 
\begin{eqnarray}
&&{\cal M}^{(m_2 \leftarrow m_1)}_{{\cal E}_n}=  
 \mbox{Tr}_{{\cal A}_n} [ {\cal V}^{(m_2 \leftarrow m_1)}_{{\cal E}_{n}{\cal A}}( \cdots \otimes |0\rangle_{{\cal A}}  \langle 0|) ]\;, \\
&& {\cal V}^{(m_2 \leftarrow m_1)}_{{\cal E}_{n}{\cal A}}(\cdots)  := V^{(m_2 \leftarrow m_1)}_{{\cal E}_{n}} (\cdots) V^{(m_2 \leftarrow m_1)^\dag}_{{\cal E}_{n}} \;, 
\end{eqnarray} 
with 
\begin{eqnarray} 
V^{(m_2 \leftarrow m_1)}_{{\cal E}_{n}} = V^{(m_2)}_{{\cal E}_{n}{\cal A}} V^{(m_2-1)}_{{\cal E}_{n}{\cal A}} \cdots V^{(m_1)}_{{\cal E}_{n}{\cal A}}\;.
\end{eqnarray} 
Hence Eq.~(\ref{impoimpo})  now rewrites as 
\begin{widetext} 
\begin{eqnarray}
2 \;  \vec{q}\;   \Omega\;  \vec{q}^{\;\dag} &=&\sum_m  \Big\langle  \hat{Q}_{{\cal E}_{n+1}}^{(m) \dag}   \hat{Q}_{{\cal E}_{n+1}}^{(m)} \cdot 
{\cal V}^{(m-1 \leftarrow 1)}_{{\cal E}_{n+1}{\cal A}}
(\hat{\eta}_{{\cal E}_{n+1}}\otimes |0\rangle_{{\cal A}}  \langle 0|
)   \Big\rangle_{\cal EA}  \nonumber \\
&&+ \sum_{m'>m}  \left[ \Big\langle
\hat{Q}_{{\cal E}_{n+1}}^{(m') \dag} \cdot 
 {\cal V}^{(m'-1 \leftarrow m)}_{{\cal E}_{n+1}{\cal A}}
\Big( \hat{Q}_{{\cal E}_{n+1}}^{(m)}\cdot 
{\cal V}^{(m-1 \leftarrow 1)}_{{\cal E}_{n+1}{\cal A}}
(\hat{\eta}_{{\cal E}_{n+1}}\otimes |0\rangle_{{\cal A}}  \langle 0|)\Big)\Big\rangle_{\mathcal{EA}} + h.c. \right]
\nonumber \\
&=& \sum_m \Big\langle \tilde{\cal V}^{(m-1 \leftarrow 1)}_{{\cal E}_{n+1}{\cal A}}( \hat{Q}_{{\cal E}_{n+1}}^{(m) \dag}   \hat{Q}_{{\cal E}_{n+1}}^{(m)} )\cdot 
(\hat{\eta}_{{\cal E}_{n+1}}\otimes |0\rangle_{{\cal A}}  \langle 0|
)   \Big\rangle_{\cal EA}  \nonumber \\
&& +\sum_{m'>m}  \left[ \Big\langle
\tilde{\cal V}^{(m-1 \leftarrow 1)}_{{\cal E}_{n+1}{\cal A}} \Big(
\tilde{\cal V}^{(m'-1 \leftarrow m)}_{{\cal E}_{n+1}{\cal A}} ( \hat{Q}_{{\cal E}_{n+1}}^{(m') \dag}) \cdot 
\hat{Q}_{{\cal E}_{n+1}}^{(m)} \Big)\cdot 
(\hat{\eta}_{{\cal E}_{n+1}}\otimes |0\rangle_{{\cal A}}  \langle 0|)\Big\rangle_{\mathcal{EA}} + h.c. \right] \;, 
 \label{impoimpo1121212} 
\end{eqnarray}
where we used the ciclicity of the trace, where $\tilde{\cal V}^{(m_2 \leftarrow m_1)}_{{\cal E}_{n}{\cal A}}$ is the conjugate transformation of ${\cal V}^{(m_2 \leftarrow m_1)}_{{\cal E}_{n}{\cal A}}$, i.e. the mapping 
\begin{eqnarray}
 \tilde{\cal V}^{(m_2 \leftarrow m_1)}_{{\cal E}_{n}{\cal A}}(\cdots)  := V^{(m_2 \leftarrow m_1)^\dag}_{{\cal E}_{n}} (\cdots) V^{(m_2 \leftarrow m_1)}_{{\cal E}_{n}} \;,
\end{eqnarray} 
and where  we   introduced the symbol ``$\cdot$" to indicate the regular product between operators whenever needed to avoid
possible misinterpretations. 
Now observe that 
\begin{eqnarray} 
\tilde{\cal V}^{(m-1 \leftarrow 1)}_{{\cal E}_{n+1}{\cal A}}( \hat{Q}_{{\cal E}_{n+1}}^{(m) \dag}   \hat{Q}_{{\cal E}_{n+1}}^{(m)} )&=&
\tilde{\cal V}^{(m-1 \leftarrow 1)}_{{\cal E}_{n+1}{\cal A}}( \hat{Q}_{{\cal E}_{n+1}}^{(m) \dag} )  \cdot  \tilde{\cal V}^{(m-1 \leftarrow 1)}_{{\cal E}_{n+1}{\cal A}}( \hat{Q}_{{\cal E}_{n+1}}^{(m)} )
=  \hat{T}_{{\cal E}_{n+1}{\cal A}}^{(m) \dag}  \; \hat{T}_{{\cal E}_{n+1}{\cal A}}^{(m) } 
\;, \\ 
\tilde{\cal V}^{(m-1 \leftarrow 1)}_{{\cal E}_{n+1}{\cal A}} \Big(
\tilde{\cal V}^{(m'-1 \leftarrow m)}_{{\cal E}_{n+1}{\cal A}} ( \hat{Q}_{{\cal E}_{n+1}}^{(m') \dag}  )\cdot 
 \hat{Q}_{{\cal E}_{n+1}}^{(m)} \Big)&=& 
\tilde{\cal V}^{(m'-1 \leftarrow 1)}_{{\cal E}_{n+1}{\cal A}} ( \hat{Q}_{{\cal E}_{n+1}}^{(m') \dag} ) \cdot 
\tilde{\cal V}^{(m-1 \leftarrow 1)}_{{\cal E}_{n+1}{\cal A}} ( \hat{Q}_{{\cal E}_{n+1}}^{(m)})= \hat{T}_{{\cal E}_{n+1}{\cal A}}^{(m') \dag}  \; \hat{T}_{{\cal E}_{n+1}{\cal A}}^{(m) } \;,
\end{eqnarray} \end{widetext} 
where we used the fact that for 
for all $m_3 > m_2 >m_1$ integer one has 
\begin{eqnarray} 
{V}^{(m_3 \leftarrow m_1)}_{{\cal E}_{n}{\cal A}} = {V}^{(m_3 \leftarrow m_2)}_{{\cal E}_{n}{\cal A}}  {V}^{(m_2 \leftarrow m_1)}_{{\cal E}_{n}{\cal A}} \;, \nonumber \\
 \end{eqnarray} 
 and introduced the operators 
 \begin{eqnarray} 
  \hat{T}_{{\cal E}_{n+1}{\cal A}}^{(m) }  = \tilde{\cal V}^{(m-1 \leftarrow 1)}_{{\cal E}_{n+1}{\cal A}} ( \hat{Q}_{{\cal E}_{n+1}}^{(m)})\;. 
  \end{eqnarray} 
 Replacing all these into Eq.~(\ref{impoimpo1121212}) and re-organizing the various terms  finally yields the thesis, i.e. 
 \begin{eqnarray}
2 \;   \vec{q}\cdot  \Omega\cdot  \vec{q}^{\;\dag}  &=&
 \sum_{m,m'=1}^M  \Big\langle \hat{T}_{{\cal E}_{n+1}{\cal A}}^{(m') \dag}  \; \hat{T}_{{\cal E}_{n+1}{\cal A}}^{(m) } 
\cdot (\hat{\eta}_{{\cal E}_{n+1}}\otimes |0\rangle_{{\cal A}}  \langle 0|
)   \Big\rangle_{\cal EA}  \nonumber \\
&\geq& 0\;. 
\end{eqnarray} 
%%%%%%%%%%%%%%
 \section{Derivation of the three body QCS master equation} \label{appendixfin} 
Here we report the explicit calculation of the model described in Fig.~\ref{fig:example2}. Following the flowchart representation presented in the right panel of the figure 
 we write the Hamiltonians~(\ref{COUPH}) as 
\begin{eqnarray}
\hat{H}_{Q_m,E_n^{(1)}}&=&\hat{a}_m^{\dagger}\hat{b}_{E_n^{(1)}}
+\hat{a}_m \hat{b}_{E_n^{(1)}}^{\dagger}\;,\\
\hat{H}_{Q_m,E_n^{(2)}}&=&0\;, 
\end{eqnarray}
where now, for $m=1,2,3$,  $\hat{a}_m$ and $\hat{a}_m^{\dagger}$ are  the lowering and raising operators of the system $Q_m$ while $\hat{b}_{E_n^{(k)}}$ and $\hat{b}_{E_n^{(k)}}^\dag$ are the bosonic operators associated with  the quantum carriers of the unidirectional channel ${\cal E}^{(k)}$ (notice that no direct coupling is assigned between 
the $Q_m$'s and ${\cal E}^{(2)}$). 
 The free dynamics of the environmental elements are instead defined by  two distinct maps:  the map $\mathcal{M}^{(1)}_{\mathcal{E}_n}(\cdots)$ associated with the beam-splitter $BS_1$ that characterizes the evolution of the quantum carriers after the interactions with $Q_1$ and before the interactions with $Q_2$; and the map $\mathcal{M}^{(2)}_{\mathcal{E}_n}(\cdots)$ associated with the beam-splitter $BS_2$ and the phase shift element $PS$ which  instead
 acts after the collisional events with $Q_2$ and before those involving  $Q_3$. Adopting the same convention used in Eqs.~\eqref{eq:mapbs1}-\eqref{eq:mapps2} they can be expressed as 
\begin{eqnarray}
\mathcal{M}^{(1)}_{\mathcal{E}_n}(\cdots)&=&\hat{V}_{BS_1}(\cdots)\hat{V}_{BS_1}^{\dagger}\;, \\
\mathcal{M}^{(2)}_{\mathcal{E}_n}(\cdots)&=&\hat{V}_{BS_2}\hat{V}_{PS}(\cdots)\hat{V}_{PS}^{\dagger}\hat{V}_{BS_2}^{\dagger}\;.
\end{eqnarray}
With this choice and assuming then the same initial conditions of Eq.~(\ref{GAUS12}) one can verify that 
stationary condition still holds for the same reasons of Sec.~\ref{sec:interferometer}, so we won't repeat the calculations of the coefficients $\gamma_{m(k)}^{(\ell)}$.
By the same token it follows that 
the local term $\mathcal{L}_1$ is identical to the one in Eq.~\eqref{LOC1ex}, because the collisional scheme is identical up to this point.
Similarly the computation of the coefficients $\gamma_{3(kk')}^{(\ell,\ell')}$, associated
with the local term of  $Q_3$, and the computation of  $\zeta_{1,3 (kk')}^{(\ell,\ell')}$ and $\xi_{1,3 (kk')}^{(\ell,\ell')}$ associated with the QCS coupling  connecting $Q_1$ with $Q_3$
coincide with the corresponding elements of $S_2$ and $S_1$ of the model of Sec.~\ref{sec:interferometer}, yielding 
the expressions reported in Eqs.~(\ref{DEFL3}) and (\ref{DEFD13}) of the main text.  
What is left is hence the computation of the terms associated with $Q_2$, i.e. ${\cal L}_2$, ${\cal D}_{1\rightarrow 2}$ and  ${\cal D}_{2\rightarrow 3}$.
Regarding the first we notice that exploiting~\eqref{eq:mapbs1} and invoking the definition  of $\bar{N}_{12}$ presetend in Eq.~(\ref{DEFNBAR12}),
the coefficients $\gamma_{2(kk')}^{(\ell,\ell')}$ can be expressed as 
\begin{eqnarray}
\nonumber
\gamma_{2(kk')}^{(1,1)}&=&[\gamma_{2(kk')}^{(2,2)}]^*=\delta_{k1}\delta_{k'1}\Big\langle \hat{b}_{E_n^{(1)}}^2\mathcal{M}_{\mathcal{E}_n}^{(1)}(\hat{\eta}_{\mathcal{E}_n})\Big\rangle=0\;, \\
\nonumber
\gamma_{2(kk')}^{(1,2)}&=&\delta_{k1}\delta_{k'1}\Big\langle \hat{b}_{E_n^{(1)}}^{\dagger}\hat{b}_{E_n^{(1)}}\mathcal{M}_{\mathcal{E}_n}^{(1)}(\hat{\eta}_{\mathcal{E}_n})\Big\rangle\\
\nonumber &=&\delta_{k1}\delta_{k'1}\Big\langle\Big(\sqrt{\epsilon_1}\hat{b}_{E_n^{(1)}}^{\dagger}+i\sqrt{1-\epsilon_1}\hat{b}_{E_n^{(2)}}^{\dagger}\Big)\\
\nonumber &\times&\Big(\sqrt{\epsilon_1}\hat{b}_{E_n^{(1)}}-i\sqrt{1-\epsilon_1}\hat{b}_{E_n^{(2)}}\Big)\hat{\eta}_{\mathcal{E}_n}\Big\rangle\\
&=&\delta_{k1}\delta_{k'1} \bar{N}_{12}, \\
\nonumber \gamma_{2(kk')}^{(2,1)}&=&\delta_{k1}\delta_{k'1}\Big\langle\hat{b}_{E_n^{(1)}}\hat{b}_{E_n^{(1)}}^{\dagger}\mathcal{M}_{\mathcal{E}_n}^{(1)}(\hat{\eta}_{\mathcal{E}_n})\Big\rangle\\
\nonumber&=&\delta_{k1}\delta_{k'1}\Big\langle\Big(\sqrt{\epsilon_1}\hat{b}_{E_n^{(1)}}-i\sqrt{1-\epsilon_1}\hat{b}_{E_n^{(2)}}\Big)\\
\nonumber&\times&\Big(\sqrt{\epsilon_1}\hat{b}_{E_n^{(1)}}^{\dagger}+i\sqrt{1-\epsilon_1}\hat{b}_{E_n^{(2)}}^{\dagger}\Big)\hat{\eta}_{\mathcal{E}_n}\Big\rangle\\
&=&\delta_{k1}\delta_{k'1} \; (\bar{N}_{12} +1 ),
\end{eqnarray}
which give  Eq.~(\ref{DEFL2new}). 
The expression  (\ref{DDEFD12}) for ${\cal D}_{1\rightarrow 2}$ instead follows from the identities 
\begin{eqnarray}
\nonumber
\zeta_{12(kk')}^{(1,1)}&=&[\xi_{12(kk')}^{(2,2)}]^*=\delta_{k1}\delta_{k'1}\Big\langle\hat{b}_{E_n^{(1)}}\mathcal{M}_{\mathcal{E}_n}^{(1)}(\hat{b}_{E_n^{(1)}}\hat{\eta}_{\mathcal{E}_n})\Big\rangle=0\;,\\
\nonumber
\zeta_{12(kk')}^{(2,2)}&=&[\xi_{12(kk')}^{(1,1)}]^*=\delta_{k1}\delta_{k'1}\Big\langle\hat{b}_{E_n^{(1)}}^{\dagger}\mathcal{M}_{\mathcal{E}_n}^{(1)}(\hat{b}_{E_n^{(1)}}^{\dagger}\hat{\eta}_{\mathcal{E}_n})\Big\rangle=0\;, \\
\nonumber \zeta_{12(kk')}^{(1,2)}&=&[\xi_{12(kk')}^{(2,1)}]^*=\delta_{k1}\delta_{k'1}\Big\langle \hat{b}_{E_n^{(1)}}^{\dagger}\mathcal{M}_{\mathcal{E}_n}^{(1)}\Big(\hat{b}_{E_n^{(1)}}\hat{\eta}_{\mathcal{E}_n}\Big)\Big\rangle\\
\nonumber&=&\delta_{k1}\delta_{k'1}\Big\langle \Big(\sqrt{\epsilon_1}\hat{b}_{E_n^{(1)}}^{\dagger}+i\sqrt{1-\epsilon_1}\hat{b}_{E_n^{(2)}}^{\dagger}\Big)\hat{b}_{E_n^{(1)}}\hat{\eta}_{\mathcal{E}_n} \Big\rangle\\
&=&\delta_{k1}\delta_{k'1}\sqrt{\epsilon_1}N_1\;,\\
\nonumber \zeta_{12(kk')}^{(2,1)}&=&[\xi_{12(kk')}^{(1,2)}]^*=\delta_{k1}\delta_{k'1}\Big\langle \hat{b}_{E_n^{(1)}}\mathcal{M}_{\mathcal{E}_n}^{(1)}\Big(\hat{b}_{E_n^{(1)}}^{\dagger}\hat{\eta}_{\mathcal{E}_n}\Big)\Big\rangle\\
\nonumber&=&\delta_{k1}\delta_{k'1}\Big\langle \Big(\sqrt{\epsilon_1}\hat{b}_{E_n^{(1)}}-i\sqrt{1-\epsilon_1}\hat{b}_{E_n^{(2)}}\Big)\hat{b}_{E_n^{(1)}}^{\dagger}\hat{\eta}_{\mathcal{E}_n} \Big\rangle\\
&=&\delta_{k1}\delta_{k'1}\sqrt{\epsilon_1}\Big(N_1+1\Big)\;,
\end{eqnarray}
while finally~(\ref{DDEFD23}) for $\mathcal{D}_{2\rightarrow3}(\cdots)$ follows from 
\begin{eqnarray}
\nonumber
\zeta_{23(kk')}^{(1,1)}&=&[\xi_{23(kk')}^{(2,2)}]^*\\
\nonumber
&=&\delta_{k1}\delta_{k'1}\Big\langle\hat{b}_{E_n^{(1)}}\mathcal{M}_{\mathcal{E}_n}^{(2)}\Big(\hat{b}_{E_n^{(1)}}\mathcal{M}_{\mathcal{E}_n}^{(1)}\hat{\eta}_{\mathcal{E}_n}\Big)\Big\rangle=0,\\
\nonumber
\zeta_{23(kk')}^{(2,2)}&=&[\xi_{23(kk')}^{(1,1)}]^*\\
\nonumber
&=&\delta_{k1}\delta_{k'1}\Big\langle\hat{b}_{E_n^{(1)}}^{\dagger}\mathcal{M}_{\mathcal{E}_n}^{(2)}\Big(\hat{b}_{E_n^{(1)}}^{\dagger}\mathcal{M}_{\mathcal{E}_n}^{(1)}\hat{\eta}_{\mathcal{E}_n}\Big)\Big\rangle=0, 
\end{eqnarray}
and
\begin{eqnarray}
\nonumber
\zeta_{23(kk')}^{(1,2)}&=&[\xi_{23(kk')}^{(2,1)}]^*\\
\nonumber
&=&\delta_{k1}\delta_{k'1}\Big\langle\hat{b}_{E_n^{(1)}}^{\dagger}\mathcal{M}_{\mathcal{E}_n}^{(2)}\Big(\hat{b}_{E_n^{(1)}}\mathcal{M}_{\mathcal{E}_n}^{(1)}\hat{\eta}_{\mathcal{E}_n}\Big)\Big\rangle\\ \nonumber 
&=&\delta_{k1}\delta_{k'1}\Big\langle\Big[c^*(\varphi)\hat{b}_{E_n^{(1)}}^{\dagger}+s^*(\varphi)\hat{b}_{E_n^{(2)}}^{\dagger}\Big]\\ \nonumber 
&&\times\Big[\sqrt{\epsilon_1}\hat{b}_{E_n^{(1)}}-i\sqrt{1-\epsilon_1}\hat{b}_{E_n^{(2)}}\Big]\hat{\eta}_{\mathcal{E}_n}\Big\rangle\\
\nonumber
&=&\delta_{k1}\delta_{k'1}\; M^*_{12}(\varphi)\;, \\
\nonumber
\zeta_{23(kk')}^{(2,1)}&=&[\xi_{23(kk')}^{(1,2)}]^*\\
\nonumber
&=&\delta_{k1}\delta_{k'1}\Big\langle\hat{b}_{E_n^{(1)}}\mathcal{M}_{\mathcal{E}_n}^{(2)}\Big(\hat{b}_{E_n^{(1)}}^{\dagger}\mathcal{M}_{\mathcal{E}_n}^{(1)}\hat{\eta}_{\mathcal{E}_n}\Big)\Big\rangle\\ \nonumber 
&=&\delta_{k1}\delta_{k'1}\Big\langle\Big[c(\varphi)\hat{b}_{E_n^{(1)}}+s(\varphi)\hat{b}_{E_n^{(2)}}\Big]\\ \nonumber 
&&\times\Big[\sqrt{\epsilon_1}\hat{b}_{E_n^{(1)}}^{\dagger}+i\sqrt{1-\epsilon_1}\hat{b}_{E_n^{(2)}}^{\dagger}\Big]\hat{\eta}_{\mathcal{E}_n}\Big\rangle\\
\nonumber
&=&\delta_{k1}\delta_{k'1} \;  (M_{12}(\varphi) + \lambda(\varphi) )  \;,
\end{eqnarray}
where we adopted the definitions~(\ref{DEFDEFDEF}). 
\begin{widetext} 
The matrix $D_{mm'(kk')}^{(\ell,\ell')}$ for this system can then be cast in the following form
\begin{eqnarray}
\nonumber
\left[\begin{array}{cc|cc|cc}
N_1 & 0 & \sqrt{\epsilon_1}N_1 & 0 &c*(\varphi)N_1 &0\\ 
0 & N_1+1 & 0 & \sqrt{\epsilon_1}(N_1+1) & 0& c(\varphi)(N_1+1)\\\hline
\sqrt{\epsilon_1}N_1 & 0 & \epsilon_1N_1+(1-\epsilon_1)N_2 & 0 & M^*_{1,2}(\varphi) & 0\\
0 & \sqrt{\epsilon_1}(N_1+1) & 0 & \epsilon_1(N_1+1)+(1-\epsilon_1)(N_2+1) & 0 & M_{1,2}(\varphi) + \lambda(\varphi) \\\hline
c(\varphi)N_1 & 0 & M_{1,2}(\varphi) & 0 & N_{12}(\varphi) & 0\\
0 & c^*(\varphi)(N_1+1) & 0 & M^*_{1,2}(\varphi) + \lambda^*(\varphi) & 0 & N_{12}(\varphi)+1
\end{array}\right]\;,
\end{eqnarray}
\end{widetext}
which upon  diagonalization yields the following  effective  Hamiltonians contributions
\begin{eqnarray}
\hat{H}_{1,2}&=&-\frac{i}{2}\sqrt{\epsilon_1}\Big(\hat{a}_1\hat{a}_2^{\dagger}-\hat{a}_1^{\dagger}\hat{a}_2\Big)\;, \label{EFFE1} \\
\hat{H}_{2,3}&=&-\frac{i}{2}\Big(\lambda^*(\varphi) \hat{a}_2\hat{a}_3^{\dagger}-\lambda(\varphi) \hat{a}_2^{\dagger}\hat{a}_3\Big)\;, \label{EFFE2} \\
\hat{H}_{1,3}&=&-\frac{i}{2}\Big(c^*(\varphi)\hat{a}_1\hat{a}_3^{\dagger}-c(\varphi)\hat{a}_1^{\dagger}\hat{a}_3\Big)\;. \label{EFFE3}
\end{eqnarray}

\end{document}